\documentclass[fleqn,usenatbib,useAMS]{mnras}

\usepackage{graphicx}	
\usepackage{amsmath}	
\usepackage{amssymb}	
\usepackage{multicol}        
\usepackage{bm}		
\usepackage{pdflscape}	
\usepackage{float}
\restylefloat{figure}
\usepackage{multirow}
\usepackage{tabularx}
\usepackage{subfig}
\usepackage{verbatim}
\usepackage{appendix}
\usepackage{fixltx2e}

\usepackage[T1]{fontenc}
\usepackage{ae,aecompl}

\usepackage{newtxtext,newtxmath}

\title[Constraining $\Lambda$CDM with Einstein Telescope]{Constraining $\Lambda$CDM cosmological parameters with Einstein Telescope mock data}

\author[M. Califano et al.]{Matteo Califano,$^{1,2}$\thanks{E-mail: matteo.califano@unina.it} Ivan de Martino,$^{3}$\thanks{E-mail: ivan.demartino@usal.es, Corresponding author} Daniele Vernieri$^{4,1,2}$\thanks{E-mail: daniele.vernieri@unina.it} and Salvatore Capozziello$^{4,1,2}$\thanks{E-mail: capozziello@unina.it}\\
	$^{1}$Scuola Superiore Meridionale, Largo San Marcellino 10, I-80138, Naples, Italy\\
	$^{2}$INFN Sezione  di Napoli, Compl. Univ. di
	Monte S. Angelo, Edificio G, Via Cinthia, I-80126, Napoli, Italy\\
	$^{3}$Universidad de Salamanca, Departamento de Fisica Fundamental, P. de la Merced S/N, Salamanca, ES\\
	$^{4}$Dipartimento di Fisica, Universit\`a
	di Napoli ``Federico II'', Compl. Univ. di
	Monte S. Angelo, Edificio G, Via Cinthia, I-80126, Napoli, Italy}

\date{Accepted XXX. Received YYY; in original form ZZZ}
\pubyear{2022}

\begin{document}
	\label{firstpage}
	\pagerange{\pageref{firstpage}--\pageref{lastpage}}
	\maketitle
	
	\begin{abstract}
		We investigate the capability of Einstein Telescope to constrain the cosmological parameters of the non-flat $\Lambda$CDM cosmological model. Two types of mock datasets are considered depending on whether or not a short Gamma-Ray Burst is detected and associated with the gravitational wave emitted by binary neutron stars merger using the THESEUS satellite. Depending on the mock dataset, two statistical estimators are applied: one assumes that the redshift is known, while the other marginalizes over it assuming a specific redshift prior distribution. We demonstrate that {\em (i)} using mock catalogs collecting gravitational wave signals emitted by binary neutron stars systems to which a short Gamma-Ray Burst has been associated, Einstein Telescope may achieve an accuracy on the cosmological parameters of $\sigma_{H_0}\approx 0.40$ km s$^{-1}$ Mpc$^{-1}$, $\sigma_{\Omega_{k,0}}\approx 0.09$, and  $\sigma_{\Omega_{\Lambda,0}}\approx 0.07$; while {\em (ii)} using mock catalogs collecting all gravitational wave signals emitted by binary neutron stars systems for which an electromagnetic counterpart has not been detected, Einstein Telescope may achieve an accuracy on the cosmological parameters of $\sigma_{H_0}\approx 0.04$ km s$^{-1}$ Mpc$^{-1}$, $\sigma_{\Omega_{k,0}}\approx 0.01$, and  $\sigma_{\Omega_{\Lambda,0}}\approx 0.01$, once the redshift probability distribution of GW events is known from population synthesis simulations and/or the measure of the tidal deformability parameter. These results show an improvement of a factor 2-75 with respect to earlier results using complementary datasets.
	\end{abstract}

	\begin{minipage}[t]{0.5\textwidth}
		\begin{flushleft}
			{\large ET preprint code: \texttt{ET-0111A-22}}
		\end{flushleft}
	\end{minipage}

	\section{Introduction}\label{sec:intro}
	
	The concordance cosmological model, known as $\Lambda$ Cold Dark Matter ($\Lambda$CDM), fits remarkably well most of the current cosmological and astrophysical observations. For instance, the {\em Planck} satellite measured with unprecedented precision the Cosmic Microwave Background (CMB) temperature anisotropies providing the most powerful tool to constrain the cosmological parameters. It favours a Universe whose accelerated expansion is determined by a cosmological constant, and whose self-gravitating structures form consequently to the gravitational collapse of baryonic matter in the potential well of the dark matter haloes~\citep{Aghanim2020b,Aghanim2020}. However, without a doubt, the three fundamental pillars of the $\Lambda$CDM model, namely inflation, dark matter, and dark energy have yet to be thoroughly tested and understood. In particular, dark matter and dark energy are necessary to obtain a good fit of cosmological and astrophysical data but their fundamental nature, {\em i.e.} whether they are particles or fields, is still a puzzle~\citep{Capozziello:2012ie, deMartino2020,Salucci2021}. 
	
	While we ignore their fundamental nature, some tensions in the $\Lambda$CDM model begin to appear as precision in observations improves~\citep{Verde2019,Abdalla2022}. The best known is the Hubble tension~\citep{Vagnozzi2020,DiValentino2021}, that is the discrepancy in the estimates of the Hubble constant obtained by local and early measurements~\citep{Philcox2020,Wong2020,Riess2021,Dainotti2021,Dainotti2022,Philcox2022}. This tension reaches the statistical level of $4.2\ \sigma$ difference and it is one of the most debated tensions in cosmology. Early measurements rely on fitting the CMB power spectra in the framework of the $\Lambda$CDM model. 
	Using Planck data, one obtains a Hubble constant equal to
	$H_0 = 67.27 \pm 0.60$ km s$^{-1}$ Mpc$^{-1}$ at 68\% of confidence level~\citep{Aghanim2020}. On the other hand, late measurements rely on {\em standard candles}, 
	which must be calibrated to build the so-called {\em cosmic distance ladder}.  Indeed, geometrical measurements of the distance obtained from the parallax are used to calibrate the luminosities of the so-called {\em standard candles}. These are astrophysical objects, such as pulsating Cepheid variables and Type Ia supernovae (SNIa), which show a tight correlation between the intrinsic luminosity and some observational quantities. Using 75 Milky Way Cepheids from the third data release of the Gaia mission, the local measurement of the Hubble constant results to be $H_0 = 73.2 \pm 1.3$ km s$^{-1}$ Mpc$^{-1}$ at 68\% of confidence level~\citep{Riess2021b}. This represents a $4.2\sigma$ with the CMB data. Although many explanations have been provided~\citep{DiValentino2021, Capozziello:2020nyq}, the debate on the origin of this tension is still open.
	
	Anyway, the Hubble tension is not the only one, there are others such as the so-called $S_8$ tension between Planck data and the weak lensing measurements, and the statistical preference of a closed Universe~\citep{DiValentino2020,Handley2021,DiValentino2021b,Nunes2021,Vagnozzi2021b,Luongo2022}. The latter appears to resolve the tension between the estimation and the expectation value of the lensing amplitude obtained from the CMB power spectra~\citep{DiValentino2020}. Nevertheless, a non-flat $\Lambda$CDM model is far from being confirmed. Indeed, the CMB temperature and polarization spectra are shown to be consistent with a spatially flat Universe~\citep{Efstathiou2020,Park2020}.
	Additionally, a recent analysis of the cosmic shear from the Kilo-Degree Survey does not find any statistical evidence in favour of a closed Universe~\citep{Troster2021}. 
	
	One way to solve these tensions lies in estimating the Hubble constant and the curvature parameter with observations complementary to the standard candles or to the CMB power spectrum. This possibility 
	relies on the Gravitational Waves (GWs) that allow us to estimate the distance of the source~\citep{Schutz1986,Capozziello2010}. GWs from a binary black hole merger were first detected in 2015~\citep{Abbott2016}. Right from the onset, GW astronomy has generated novel ways to test fundamental physics: clear examples are the tests of gravity in the strong-field regime~\citep{Abbott2016, Abbott2016b}, and the first high-precision measurement of the GW speed~\citep{Abbott2017} which spectacularly ruled out large classes of gravitational theories~\citep{Ezquiaga2017}.  The opportunities to test the dark Universe will multiply as GW data continues to improve and new experiments come online, {\em e.g.} Einstein Telescope (ET)~\citep{Maggiore2020}. Nevertheless, GWs are not free of issues. The luminosity distance can be determined from the waveform, otherwise, we cannot estimate the redshift because it is degenerate with the chirp mass. This well-known degeneracy can be broken by obtaining the redshift information from an electromagnetic counterpart such as the emission of a short Gamma-Ray Burst (GRB)~\citep{Capozziello2011,Abbott2017} or the optical and spectroscopic localization of the host galaxy~\citep{Holz2005,Chen2018} though the latter will not be accurate for high redshift events~\citep{Diaz:2022}. Another solution, that has been recently proposed, relies on obtaining the redshift from the tidal deformation in a Binary Neutron Stars (BNS) merger~\citep{messenger:Read,Chatterjee2021}. This solution may supply an estimation for the redshift with an accuracy ranging from  $8$\% to $40$ \%, depending on the choice of the equation of state. 
	
	Here, we want to study the following issues: first, we want to forecast the precision down to which ET will be able to constrain the Hubble constant $H_0$ and the curvature parameter $\Omega_{k,0}$ and its capability to resolve the aforementioned tensions using BNS signals. Several analyses have been carried out to measure the Hubble constant with ET~\citep{Zhao:2010sz,Sathyaprakash2010,Cai:2017,D'Agostino2019,Yang2019,Bonilla2020,Belgacem:2019tbw,Zhang2019,Ssohrab:2020,Yu2021,Jin2021,Yang:2021qge,Jin2022,Herbett:2022,jin2022b, Alfradique2022}, but they were considering only a flat-$\Lambda$CDM model. Second, since ET will detect more than 20,000 BNS events/year at Signal-to-Noise ratio (SNR) $\geq 9$ but only for $\sim 0.1\%$ of them the electromagnetic counterpart can be observed (as we will show in Sect. \ref{sec:three}), we want to analyse alternatives to the direct observation of the electromagnetic counterpart to investigate whether they can represent a viable way to achieve a better precision on the cosmological parameters. The manuscript is sectioned as follows: we first briefly introduce the fiducial cosmological model used to build the mock catalogs in Section~\ref{sec:two}. The details on the procedure used to build them are given in Section~\ref{sec:three}. In Section~\ref{sec:four}, we give the details of our statistical analysis used to forecast the precision of cosmological parameters achievable with ET, and in Section~\ref{sec:five} we show all results. Finally, in Section~\ref{sec:six}, we summarize our results and give our main conclusions.
	
	\section{The $\Lambda$CDM cosmological model}\label{sec:two}
	
	Let us start by briefly summarizing the $\Lambda$CDM cosmological model which is based on the Friedmann--Lemaitre--Robertson--Walker (FLRW) metric
	\begin{equation}\label{FLRWmetric}
		ds^2=-c^2dt^2 + a^2(t)\left[ \frac{dr^2}{1-kr^2}+ r^2\left(d\theta^2 + \sin{\theta}^2 d\phi^2 \right) \right],
	\end{equation}
	where $c$ is the speed of light, $t$ is the cosmic time, $a(t)$ is the scale factor whose evolution depends on the matter and energy contents of the Universe, and $k = +1,-1, 0$ corresponds to a closed, open, and flat universe, respectively.
	From the Einstein equations, one can derive the Friedmann equations
	\begin{align}
		H^2(t) &= \frac{8 \pi G}{3}\rho(t) -\frac{kc^2}{a^2(t)},\label{Friedeq}\\
		\frac{\ddot{a}(t)}{a(t)}&=-\frac{4\pi G}{3}\left(\rho(t) + \frac{3p(t)}{c^2}\right) \label{Friedeq2}\,,
	\end{align}
	where $G$ is the gravitational constant, $H(t)$ is the Hubble parameter and $\rho(t)$ is the energy-matter density given by $\rho(t)=\rho_{r}(t) + \rho_{m}(t) + \rho_{\Lambda}$. The first term represents the density of radiation, the second term represents the matter density and the last term represents the cosmological constant density. Then, defining the redshift $z$ in terms of scale factor and using the definition of the critical energy density, $\rho_c = \frac{3H^2}{8\pi G}$, Eq.~\eqref{Friedeq} can be recast as follows:
	\begin{equation} \label{sum:density}
		1 = \Omega_{r}(z) + \Omega_{m}(z) + \Omega_{\Lambda} + \Omega_{k}(z),
	\end{equation}
	where the following definitions $\Omega_{i}(z)\equiv\frac{\rho_i(z)}{\rho_c(z)}$ and $\Omega_{k}(z)=\frac{-kc^2(1+z)^2}{ H^{2}(z)}$ have been introduced.
	Finally, the Eq.~\eqref{Friedeq} can be also recast as
	\begin{equation}\label{eq:Ez}
		\frac{H^2(z)}{H_{0}^{2}} \equiv E^2 (z)  = \Omega_{r,0}(1+z)^{4} + \Omega_{m,0}(1+z)^{3} + \Omega_{k,0}(1+z)^{2} + \Omega_{\Lambda,0}\,,
	\end{equation}
	which is required to compute the comoving distance  
	\begin{equation}\label{comoving_dist}
		d_C(z) = D_H \int_0^z \dfrac{1}{E(z)} dz\ ,
	\end{equation}
	where $D_H = \frac{c}{H_0}$ is the Hubble distance.
	Consequently, the luminosity distance is~\citep{Hogg:1999ad}:
	\begin{equation}\label{luminosity_distance}
		d_L(z) = S_k\biggl(\sqrt{|\Omega_{k,0}|}\frac{d_C}{D_H}\biggr)(1+z)\ ,
	\end{equation}
	where
	\begin{equation}
		S_k\biggl(\sqrt{|\Omega_{k,0}|}\frac{d_c}{D_H}\biggr)  = 
		\begin{cases}
			\frac{d_H}{\sqrt{|\Omega_{k,0}|}} \sinh \biggl(\sqrt{|\Omega_{k,0}|}\frac{d_c}{D_H}\biggr) &\text{if $\Omega_{k,0} > 0$}\\
			d_c &\text{if $\Omega_{k,0} = 0$}\\
			\frac{d_H}{\sqrt{|\Omega_{k,0}|}}\sin \biggl(\sqrt{|\Omega_{k,0}|}\frac{d_c}{D_H}\biggr) &\text{if $\Omega_{k,0} < 0$}\end{cases}
	\end{equation}
	
	To build the mock catalogs of cosmological distances, we set as a {\em fiducial} cosmological model the $\Lambda$CDM with the following observational constraints:
	\begin{equation}\label{fid_cosmo}
		H_0 = 67.66\ \mbox{km}\ \mbox{s}^{-1}\mbox{Mpc}^{-1},\ \Omega_{k,0}=0.0,\ \Omega_{\Lambda,0}= 0.6889\,.
	\end{equation}
	
	Since $\Omega_{r,0}\sim 10^{-5}$~\citep{Fixsen2009}, its contribution to the cosmic evolution of the Universe is negligible, in the range of redshift we will explore, compared the other components. Therefore, we set $\Omega_{r,0}=0$. Finally, the contribution of the matter density, $\Omega_{m,0}$, is derived using Eq.~\eqref{sum:density} as $\Omega_{m,0}=1 -( \Omega_{\Lambda,0} + \Omega_{k,0})$.

	\section{Construction of mock source catalogs}\label{sec:three}
	
	Let us start by giving a schematic description of the main ingredients needed to build the mock catalogs:
	\begin{description}
		\item[{\em (i)}] the first step is to define the probability distribution $p(z)$ of an event happening at redshift $z$. This probability distribution will depend on the astrophysics, {\em i.e.} the merger rate;
		\item[{\em (ii)}] the merger rate of BNS defines the rate density per unit of redshift in the observer frame, and it depends on the Star Formation Rate (SFR);
		\item[{\em (iii)}] one needs to define the mass distribution of the NS. We set it to be uniform in the interval $[1, 2.5]\ M_\odot$;
		\item[{\em (iv)}] one needs to set the spatial distribution of the merger event of BNS. We set it to be isotropic on the sky angles $\theta$ and $\phi$, and uniform on the orientation angle $\cos i$ and the polarization $\psi$.
	\end{description}
	
	Once the previous ingredients have been defined, and using the {\em fiducial} cosmological model introduced in Section~\ref{sec:two}, we can extract the {\em fiducial} redshift of the GW source from the probability distribution $p(z)$. Then, we can predict the SNR, $\rho$, for ET using the expected one-sided noise power spectral density. Finally,  we select events having SNR above a fixed threshold. Specifically, we will create three mock catalogs with SNR thresholds equal to $[9, 12, 15]$ for each observational period fixed in one, five, and ten years. In total, we have nine mock catalogs containing all events that ET will be capable of detecting. From each one, we extract two sub-catalogs listing only the events of BNS merger that have a detected electromagnetic counterpart. Hence from each of the nine initial catalogs we finally obtain three catalogs, for a total of 27 mock catalogs.

	\subsection{Probability distribution and  merger rate of the GW events}
	The normalized probability distribution of GW events is defined as follows~\citep{RegimbauHughes:2009,Regimbau:2012,Regimbau:2014,Meacher:2016,Cai:2017,Regimbau:2017,Belgacem:2019tbw}:
	\begin{equation}\label{red_distr}
		p(z) = \frac{R_z (z)}{\int_{0}^{z_{max}}R_z (z)dz}.
	\end{equation}
	
	Here, we set the maximum observed redshift $z_{max}=10$~\citep{Belgacem:2019tbw}.  $R_z(z)$ is the merger rate density per unit of redshift in the observer frame
	\begin{equation}\label{rate:unit_of_redshift}
		R_z(z) = \frac{R_m (z)}{1+z}\frac{dV(z)}{dz},
	\end{equation} 
	where $dV(z)/dz$ is the comoving volume element, and $R_m (z)$ is the merger rate per volume in the source frame. The comoving volume can be written in terms of the comoving distance given in Eq.~\eqref{comoving_dist} as
	\begin{equation}
		\frac{dV(z)}{dz} = \frac{4\pi D_H d_C^2(z)}{E(z, \Omega_{k,0}, \Omega_{\Lambda,0})}.
	\end{equation}
	On the other side, the merger rate per unit of volume at redshift $z$ is related to the SFR  through the time delay distribution $P(t_d)$. The latter is the probability distribution of the time delay, $t_d$, between the time at the formation of the massive progenitors ($t_f$) and the time at which they merge ($t_m$), \textit{i.e.} $t_d= t_f - t_m$.  The time  delay distribution can be set to two different functional forms: {\em (i) power law} form $P(t_d)\propto t_{d}^{-1}$ as suggested by population synthesis models~\citep{Tutukov:1994,Lipunov:1995,Pacheco:2006,Belczynski:2006,Shaughnessy:2008}, or  {\em (ii) exponential} form $P(t_d)\propto \tau^{-1}\exp(-t_d / \tau)$  as suggested by~\cite{Vitale:2020}, with an e-fold time of $\tau=100\ \mbox{Myr}$ for the time delay distribution. Finally, the merger rate takes the following form
	\begin{equation}\label{merger_rate}
		R^{*}_{m} (z) = \int_{t_{min}}^{t_{max}} R_f[t(z)-t_d] P(t_d) d t_d \ ,
	\end{equation}
	where $t_{min}=20$ Myr and $t_{max}$ is the Hubble time~\citep{Meacher:2015iua}, $R_f(t)$ is the formation rate of massive binaries and $t(z)$ is the age of the Universe at the time of merger.  Additionally, the formation rate of massive binaries follows the model for SFR~\citep{RegimbauHughes:2009}, that can be set to the {\em Vangioni model}~\citep{Vangioni:2014axa}
	\begin{equation}\label{VAN:SFR}
		\mathrm{SFR}(z)= \frac{\nu \ a \exp(b(z-z_m))}{a-b+b \exp(a(z-z_m))},
	\end{equation}
	with parameters $a= 2.76$, $b = 2.56$, $z_m = 1.9$ and $\nu=0.16\ M_{\odot}\mbox{/yr/Mpc}^3$. Alternatively, the SFR can be also set to {\em Madau--Dickinson model}~\citep{Madau:2014bja}
	\begin{equation}\label{MD:SFR}
		\mathrm{SFR}(z)= \frac{(1+z)^{\alpha}}{1+\left[ (1+z)/C\right]^{\beta}},
	\end{equation}
	with $\alpha = 2.7$, $\beta = 5.6$ and $C=2.9$.
	
	Finally, we normalize Eq.~\eqref{merger_rate} by requiring that the value of $R_m (z=0)$ agrees with the most recent estimation of the local rate made by~\citet{LIGOScientific:2021psn}:  $R_{m}(z=0)=105.5^{+190.2}_{-83.9}\ \mbox{Gpc}^{-3} \mbox{yr}^{-1}$. Their estimation is inferred assuming
	that neutron stars masses range from 1 to 2.5 $M_{\odot}$. 
	Thus, the merger rate can be recast as:
	\begin{equation}\label{merg_rate_final}
		R_{m}(z)= R_m(z=0) \frac{R^{*}_{m}(z)}{R^{*}_{m}(z=0)}.
	\end{equation}
	
\begin{figure}
		\centering
		\subfloat{\includegraphics[width=0.48\textwidth]{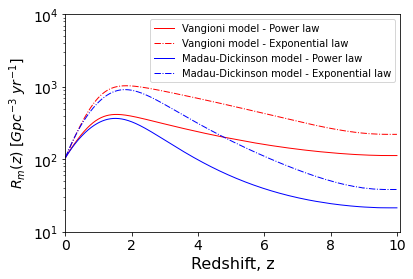}}
		\quad
		\subfloat{\includegraphics[width=0.48\textwidth]{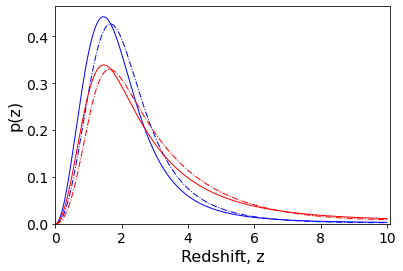}}
		\caption{\textit{Top:} The BNS merger rate as function of the redshift.\textit{Bottom:} The normalized probability distribution of redshift.  In both panels, we depict the Madau - Dickison model given in Eq.~\eqref{MD:SFR} (blue lines) and the Vangioni model given in Eq.~\eqref{VAN:SFR}  (red line). Furthermore, we show both a power law and an exponential time delay distribution as  solid  and dashed lines, respectively. In our {\em baseline} model, we adopt the Vangioni model for SFR and the power law time delay distribution (red solid line). }
		\label{fig:pdf_BNS}
	\end{figure}

	
	In Figure~\ref{fig:pdf_BNS}, we show the normalized probability distribution of redshift (bottom panel) and the normalized BNS merger rate density (top panel) for different assumptions for SFR and time delay distribution.
	Once the probability distribution of GW events is built, we can extract the redshift of the GW events according to this distribution in framework of our {\em baseline} model, which adopts the {\em Vangioni model} for the SFR and the {\em power law} form of the time delay distribution. We will show in Section~\ref{sec:check} that making other choices leads to similar constraints on the cosmological parameters.

	\subsection{Detector sensitivity}
	
	The next step is to compute the SNR of the event detected by the ET. The observatory will have three independent interferometers and, hence, the combined SNR  is~\citep{Finn2001,Schutz2011}:
	\begin{equation}
		\rho=\sqrt{\sum\limits_{i=1}^{3}(\rho^{(i)})^2}.
	\end{equation}
	
	Under the assumption of Gaussian noise, the $i-$th interferometer will have a SNR~\citep{FinnChernoff:1993}
	\begin{equation}
		\rho^{2}_{i}=4 \int _{0}^{\infty} \frac{|\tilde{h}(f)|}{S_{h,i}(f)} df,
	\end{equation}
	where $f$ is the GW frequency in the observer frame, $S_{h,i}(f)$ is the one-sided noise power spectral density of $i$-th interferometer, and $\tilde{h}$ is the Fourier transform of the GW strain amplitude~\citep{Blanchet:1996}:
	$\tilde{h}= F_{+,i}\tilde{h}_{+}+F_{\times,i}\tilde{h}_{\times}$; where
	$\tilde{h}_{+}$ and $\tilde{h}_{\times}$  are the GW strain amplitudes of $+$ and $\times$ polarizations, and $F_{+,i}(\psi,\theta,\phi)$ and $F_{\times,i}(\psi,\theta,\phi)$ are the so-called beam pattern functions \citep{Zhao:2010sz}
	\begin{equation}
		\begin{aligned}
			F_+^{(1)}(\theta, \phi, \psi)=&\frac{{\sqrt 3 }}{4}\left[(1 + {\cos ^2}(\theta ))\cos (2\phi )\cos (2\psi )\right.\\
			&\left.\qquad- 2\cos (\theta )\sin (2\phi )\sin (2\psi )\right],\\
			F_\times^{(1)}(\theta, \phi, \psi)=&\frac{{\sqrt 3 }}{4}\left[(1 + {\cos ^2}(\theta ))\cos (2\phi )\sin (2\psi )\right.\\
			&\left.\qquad +2\cos (\theta )\sin (2\phi )\cos (2\psi )\right],\\
			F_{+,\times}^{(2)}(\theta, \phi, \psi)=&F_{+,\times}^{(1)}(\theta, \phi+2\pi/3, \psi),\\
			F_{+,\times}^{(3)}(\theta, \phi, \psi)=&F_{+,\times}^{(1)}(\theta, \phi+4\pi/3, \psi),
		\end{aligned}
	\end{equation} 
	where $\phi$ and $\theta$ are the sky localization angles, and $\psi$ is the polarization angle. We select $\phi$ and $\psi$ angles randomly from a uniform distribution in the range $[0, 2\pi]$, whereas $\theta$ belongs to the range $[0, \pi]$.
	At quadrupolar order, the GW strain amplitudes of the polarization states can be written as~\citep{FinnChernoff:1993}
	\begin{align}
		\tilde{h}_{+}=&\frac{1}{\pi^{2/3}}\left(\frac{5}{24}\right)^{\frac{1}{2}}\frac{c}{d_L}\left(\frac{G\mathcal{M}_{c}}{c^{3}}\right)^{\frac{5}{6}} f^{-\frac{7}{6}}\left(\frac{1+\cos{i}^{2}}{2} \right),\\
		\tilde{h}_{\times}=&\frac{1}{\pi^{2/3}}\left(\frac{5}{24}\right)^{\frac{1}{2}}\frac{c}{d_L}\left(\frac{G\mathcal{M}_{c}}{c^{3}}\right)^{\frac{5}{6}} f^{-\frac{7}{6}}\cos{i},
	\end{align}
	where $\mathcal{M}_{c}$ is the observed chirp mass defined as a combination of individual masses $m_1$ and $m_2$,
	$\mathcal{M}_{c}=(1+z) \frac{(m_1 m_2)^{3/5}}{(m_1 + m_2)^{1/5}} $.
	Since the dominant contribution to SNR in a BNS event comes from the inspiral phase~\citep{FinnChernoff:1993}, the SNR can be written as
	\begin{equation}\label{SNR}
		\rho^{2}_{i}=\frac{5}{6}\frac{(G \mathcal{M}_{c,obs})^{\frac{5}{3}}\mathcal{F}_{i}^{2}}{c^{3}\pi^{\frac{4}{3}}d_{L}^{2}(z)}\int_{f_{\rm lower}}^{f_{\rm upper}}\frac{f^{-\frac{7}{3}}}{S_{h,i}(f)}df,
	\end{equation}
	where 
	\begin{equation}
		\mathcal{F}_{i}^{2}=\left(\frac{1+\cos^{2}{i}}{2} \right)^{2}F_{+,i}^{2} + \cos^{2}{i}F_{x,i}^{2},
	\end{equation}
	and $\cos^{2}{i}$ is the cosine square of the inclination angle. The latter is also selected randomly from a uniform distribution ranging from zero to $\pi$.
	The upper limit of the integration, $f_{upper}$, is related to the frequency of the  last stable orbit (LSO), $f_{LSO}$, which marks the end of the inspiral phase and the onset of the final merger~\citep{Maggiore:2007}
	\begin{equation}
		f_{upper}=2f_{LSO}=\frac{c^3}{(6\sqrt{6}\pi M_{obs})G},
	\end{equation}
	where
	$M_{obs} = (1+z)\,(m_1+m_2)$  is the observed total mass in the reference frame of the observer.
	The lower cutoff of the integration, $f_{lower}$, depends on the detector.  In case of the ET the lower limit frequency is $f_{lower}=1$ Hz~\citep{Sensitivity:2011}. Finally, $S_h(f)$ is the latest sensitivity curve made available by the ET collaboration\footnote{The latest power spectral density $S_h(f)$ can be downloaded at \url{https://apps.et-gw.eu/tds/?content=3&r=14065}.}.
	Once the SNR is computed, we can build our mock catalogs by selecting all the events above a given threshold.
	
	\subsection{The mock catalogs}
	
	The detection rate of sources depends on the sensitivity, observing time, and duty cycle of the GW detector network, {\em i.e.}, the time percentage of the detectors in science mode~\citep{RegimbauSiellez:2015}. We set the duty-cycle of the ET to $80\%$ as in Ref.~\citep{RegimbauSiellez:2015}. In other words, ET will detect the  $80\%$ of the total number of events. The latter is estimated integrating the Eq.~\eqref{rate:unit_of_redshift} over the redshift interval $z=[0,10]$ and multiplying it by  1, 5 or 10 years of observations. Additionally, only events above a given threshold are retained. More specifically, we set three different thresholds in SNR equal to 9, 12, and 15. Finally, we add a Gaussian noise component, $\mathcal{N}(d_{L}^{fid},\sigma_{ d_{L}})$, to our estimations of the luminosity distances $d_{L}^{fid}$ based on the {\em fiducial} cosmological model. 
	The variance $\sigma_{ d_{L}}$ accounts for different sources of uncertainties:
	\begin{equation}\label{sigma_dl}
		\sigma_{d_L}=\sqrt{\sigma_{inst}^2+\sigma_{delens}^2 +\sigma_{pec}^2}\,. 
	\end{equation}
	The first term $\sigma_{inst}$ is related to the instrumental error and is proportional to $1/\rho$~\citep{Cutler1994,Dalal:2006}:
	\begin{equation}
		\sigma_{inst}=\frac{2}{\rho}d_L(z).
	\end{equation}
	The factor two accounts for the degeneration between $\rho$ and the inclination angle, which may differ for each event.
	However, the detection of the  electromagnetic counterpart breaks the degeneracy between distance and inclination angle and improves the accuracy by a factor of two~\citep{Nissanke:2009kt}. Anyway, to be conservative, we retained the factor two in $\sigma_{inst}$.
	
	The second contribution  to the total uncertainty, $\sigma_{lens}$,  accounts for the weak lensing distorsions~\citep{Hirata:2010,Tamanini:2016}
	\begin{equation}
		\sigma_{lens}=0.066\left(\frac{1-(1+z)^{-0.25}}{0.25}\right)^{1.8}d_L(z).
	\end{equation}
	
	Nevertheless, the future detectors such as the Extremely Large Telescope will estimate  the lensing magnification distribution along the line of sight of the GW event~\citep{Shapiro2010}, and they will allow to reduce the uncertainty due to weak lensing by applying  a de-lensing factor $F_{delens}(z)$~\citep{Speri:2021}
	\begin{equation}
		\sigma_{delens}=F_{delens}(z) \left( \sigma_{ d_{L}} (z)\right)_{lens},
	\end{equation}
	where $F_{delens}(z)= 1- \frac{0.3}{\pi /2}\arctan{\frac{z}{z_*}}$, with $z_*=0.073$.
	
	The last contribution to the total uncertainty, $\sigma_{pec}$, is related to the peculiar velocities, and can be estimated by mean of a fitting formula~\citep{Kocsis:2006}
	\begin{equation}
		\sigma_{pec}=\left[ 1+\frac{c(1+z)^2}{H(z)d_L (z)}\right]\frac{\sqrt{\langle v^2\rangle}}{c}d_L (z)\,,
	\end{equation}
	where we set the averaged peculiar velocity $\langle v^2\rangle$ to  $500$ km/s, in agreement with the observed values in galaxy catalogs~\citep{Cen2000}. 
	\begin{table}
		\centering
		\caption{Numbers of GW events from BNS merger recorded in each mock catalog depending on the threshold in the SNR and the total number of observational years.}
		\label{Tab:events}
		\begin{tabular}{c|c|c||c|c||c|c}
			\hline
			\multirow{2}{*}{\bf{Years}}&\multicolumn{2}{c||}{\bf SNR = 9}  & \multicolumn{2}{c||}{ {\bf SNR = 12}} & \multicolumn{2}{c}{\bf SNR = 15}\\
			\cline{2-7}
			&  {\bf \# events} & ${\boldmath z_{max}} $ &  {\bf  \# events} & ${\boldmath z_{max}}$  &  {\bf  \# events} & ${\boldmath z_{max}}$ \\
			\hline
			1 & 52309 & 6.23 & 29594 & 3.97 & 17235 & 2.65 \\
			5 & 260088 & 6.77 & 146512 & 4.03 & 85595 & 2.86 \\
			10 & 521552 & 6.91 & 294284 & 4.05  & 171270 & 2.86 \\
			\hline
		\end{tabular}
	\end{table}
	
	\begin{figure*}
		\centering
		\includegraphics[width=0.98\textwidth]{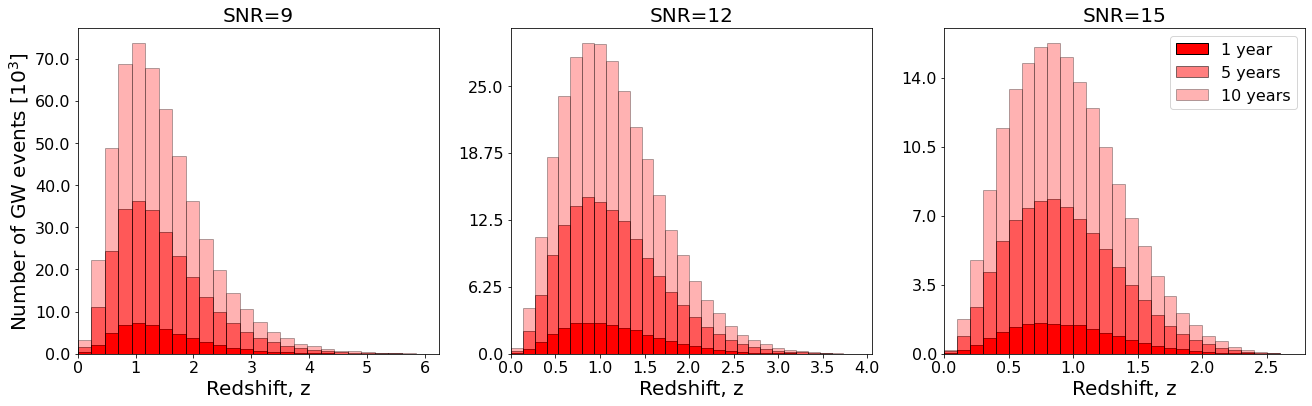}
		\caption{Distribution of the expected rate of detection of GW events from BNS merger with ET. Left, central, and right panels illustrate the detection rate for SNR$>[9, 12, 15]$, respectively.}
		\label{fig:event_distr}
	\end{figure*}

	In Table~\ref{Tab:events}, we report the number of events and the maximum observed redshift corresponding to each catalog. We  built a total of nine catalogs corresponding to the one, five and ten years of observing mode, and to SNR equal to 9, 12, and 15. The rate of events that we have obtained is comparable with the one expected for ET, as shown in~\citet{Maggiore2020}. In Figure~\ref{fig:event_distr}, we illustrate the observed events distribution as function of the redshift. The left, central, and right panels correspond to the SNR thresholds 9, 12, and 15, respectively.
	Each panel depicts the redshift distribution obtained after one, five and ten years of observations with dark red, red, and light red colors as shown in the legend.
	
	\subsection{Detection of the electromagnetic counterpart with THESEUS}
	
	On August 17, 2017 at $12$:$41$:$04$ UTC a new era of gravitational and multi-messenger astronomy began.
	The Advanced LIGO (aLIGO) and Advanced Virgo (aVIRGO)
	GW detectors observed the first merger of a BNS, GW170817, with the simultaneous detection of the Gamma-Ray Burst (GRB) 170817A associated with a GW event~\citep{Abbott2017d}.  This event was extremely important not only to constrain astrophysical scenarios for the production of a GRB~\citep{Abbott2017d}, but also to constrain fundamental physics and modified gravity theories. For instance, the tiny time lag detected between the GWs and their electromagnetic counterpart served to rule out several modified theories of gravity where GWs and photons travel at two different speeds, and to reduce the parameter space of the Effective Field Theory of Dark Energy 
	\citep{Baker2017,Creminelli2017,Ezquiaga2017,Sakstein2017,Wang2017,Wei2017,Amendola2018,Cai2018,Crisostomi2018,Kreisch2018}. Hence, it results extremely important to known how many events that will be detected by ET, will also be detected by forthcoming Gamma-Ray and X-Ray satellites. In the following, we will focus on the Transient High Energy Sources and Early Universe Surveyor (THESEUS) satellite that could overlap with ET and provide the electromagnetic counterpart of the GW events~\citep{THESEUS:2017qvx,THESEUS:2017wvz,Amati2021,Ciolfi2021,Ghirlanda2021,Rosati2021}.

	Once the GW catalogs have been built, we look for those events with a detectable electromagnetic counterpart in order to obtain the corresponding redshift of the event and  break the mass-redshift degeneracy~\citep{FinnChernoff:1993} to use the binary NS as standard (bright) sirens. As mentioned above, we focus on the observational specifications of the THESEUS satellite and, following~\citet{Yang:2021qge}, we simulate the observed photon flux of the GRB events associated with a GW event through the luminosity distance by sampling the luminosity  probability distribution $\phi(L)$. Then, if the observed flux is above the given threshold, we retain the event. There are two ways of identifying  the source of the electromagnetic counterpart in the sky. On one side, one can localize the source and follow-up it  with optical telescopes, as done to detect  the electromagnetic counterpart of the event GW170817.  ET will localize the most of the detectable BNS at a distance $ \leq$ 200 Mpc to within a 90$\%$ credible region of 100 deg$^2$~\citep{Chan:2018csa,Zhao:skyloc,Li:2021mbo}. 
	On the other side, even without a precise sky localization,  joint observations are possible. Indeed, the time coincidence of the GW event with a short GRB can be used and then the redshift can be determined from the X-ray afterglow.

	Starting from the mock catalogs of the GW events, we want to extract those events having a detected electromagnetic counterpart. To this aim, we need to know observational features  of the short GRBs such as the viewing angle and the sharpness of the angular profile. Thus,  we assume the following luminosity profile of the jet~\citep{Resmi2018}
	\begin{equation}\label{jet_profile}
		L(\theta_{V})= L_c e^{-\frac{\theta_{V}^{2}}{2\theta_{c}^2}}\,,
	\end{equation}
	where $L(\theta)$ is the luminosity per unit of solid angle, $\theta_{V}$ is the viewing angle and $L_{c}$ and $\theta_{c}$ are the structure parameters that define the sharpness of the angular profile. A jet with a large $\theta_{c}$ is similar to a uniform jet. Following the analysis made by~\citet{Howell:2019}, we set $\theta_{c} = 4.7^{\circ}$. 
	The value of $L_c$ is related to the peak luminosity of each burst, $L_p$, by the relation $L_c = \frac{L_p}{4\pi}\ \mbox{erg}\ \mbox{s}^{-1} \mbox{sr}^{-1}$~\citep{Resmi2018}.
	We then assume that $L_p$ is distributed following a standard broken power law~\citep{Wanderman:2014eza}
	\begin{equation}\label{distr_Lp}
		\Phi(L_p) \propto
		\begin{cases}
			\left(\dfrac{L_p}{L_{*}}\right)^{\alpha}\,\, \quad (L_{min}<L_p<L_{*}),\\[0.4cm]
			\left(\dfrac{L_p}{L_{*}}\right)^{\beta}\,\, \quad (L_p>L_{*}),\end{cases}
	\end{equation}
	where $L_p$ is the isotropic rest frame luminosity in the $1-10^4$ keV energy range, and $L_{*}$ represents a characteristic luminosity that separates the low and high regimes of the luminosity function  encoded in  the value of the coefficients $\alpha$ and $\beta$. Finally, $L_{min}$ is the low end cutoff.
	We set these parameters to the values given by~\citet{Wanderman:2014eza}
	$ \alpha= -1.95$, $\beta = -3$,$ L_{*} = 2 \times10^{52} \mbox{erg}\ \mbox{s}^{-1}$, and  $L_{min} = 10^{49}\ \mbox{erg}\ \mbox{s}^{-1}$.
	
	Next, by sampling the luminosity distribution function in Eq.~\eqref{distr_Lp}, we get a value of $L_p$ for each burst that must be  converted to an observed flux. 
	However, the relation flux-luminosity requires two correction factors~\citep{Meszaros11}:
	\begin{equation}\label{eq:flux}
		F(\theta_{V})=\frac{\left(4\pi L(\theta_{V})\right)(1+z)}{4\pi d_{L}^{2}\ k(z)\ b}.
	\end{equation}
	The first correction term, $b$, is an energy normalization due to scaling of observed photon flux to account for the missing fraction of the gamma-ray energy seen in the detector band:
	\begin{equation}
		b=\frac{\int_{1\ keV}^{10000\ keV}EN(E)dE}{\int_{E1}^{E2}N(E)dE},
	\end{equation}
	where $E1$, $E2$ are the extreme of the detector’s energy window,  and $N(E)$ is the band function. 
	The second correction term, $k(z)$, is due to redshifted photon energy, also called $k$-correction, and it is given by
	\begin{equation}
		k(z)=\frac{\int_{E1}^{E2}N(E)dE}{\int_{E1(1+z)}^{E2(1+z)}N(E)dE}.
	\end{equation}
	The band function $N(E)$ is given by a phenomenological fit to the observed GRB photon spectrum~\citep{Band:1993}
	\begin{equation}
		N(E)=
		\begin{cases}
			N_0 \left( \frac{E}{100 \mbox{keV}} \right)^{\alpha_{B}}\exp{-\frac{E}{E0}} \ \ \ \qquad \qquad \quad \qquad \qquad (E < E_b),\\[0.4cm]
			N_0 \left( \frac{E_b}{100 \mbox{keV}} \right)^{\alpha_{B}-\beta_{B}}\exp{\left(\beta_{B}-\alpha_{B}\right)}\left( \frac{E}{100 keV} \right)^{\beta_{B}} \ \ (E >E_b),\end{cases}
	\end{equation}
	where $E_b = (\alpha_B - \beta_B)E_0$ and $E_p = (\alpha_B + 2)E_0$.
	We set $\alpha_B = -0.5$, $\beta_B = -2.25$ and a peak energy $E_p = 800\ \mbox{keV}$ in the source frame as in~\citep{NAVA:2011}. 
	Finally, the factor $(1+z)$ in the Eq.~\eqref{eq:flux} is related to the fact that the flux is measured in units of $\left[\frac{\mbox{photon}}{\mbox{cm}^2\ \mbox{s}}\right]$~\citep{Meszaros11}.
	
\begin{figure*}
		\centering
		\includegraphics[width=0.98\textwidth]{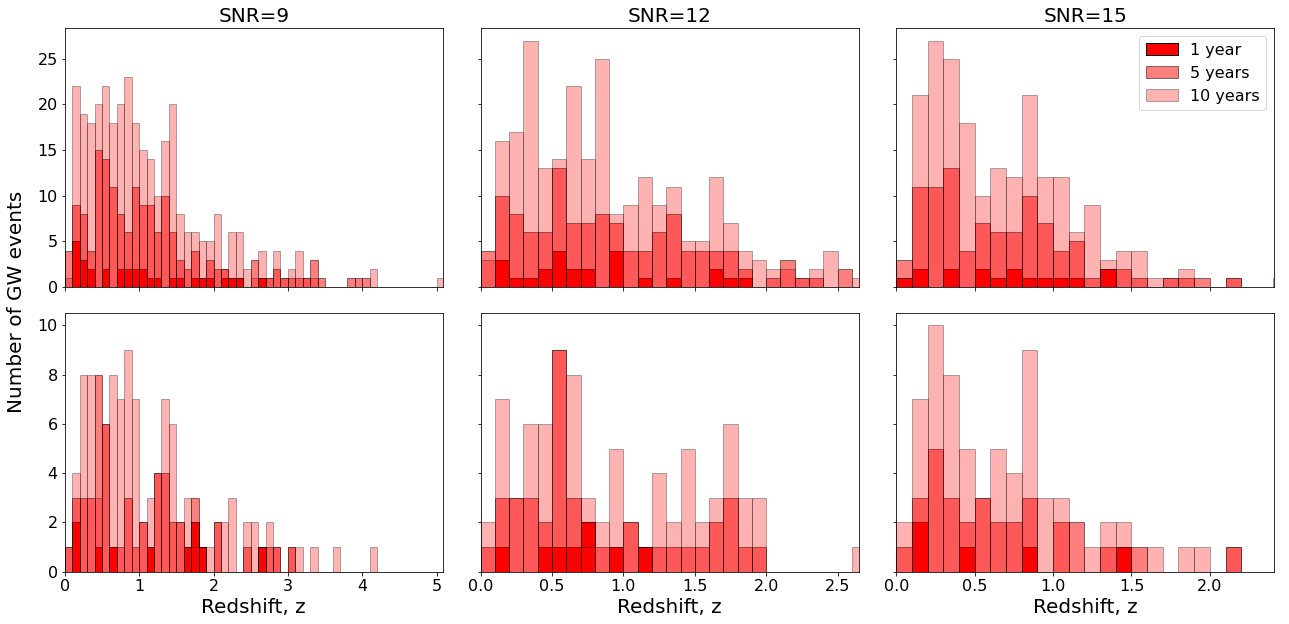}
		\caption{Distribution of the expected rate of detection of GW events with an electromagnetic counterpart for both the {\em optimistic} and the {\em realistic} cases in upper and lower row, respectively. Left, central, and right panels illustrate the detection rate for SNR$>[9, 12, 15]$, respectively.}
		\label{fig:event_distr2}
	\end{figure*}

	THESEUS satellite  will detect a 15 - 35 coincident short GRB  per year~\citep{THESEUS:2017wvz}, which represent only the $\sim0.08\%$ of the expected rate of GW events detected by ET at SNR$>9$ (see Table~\ref{Tab:events}). We set the duty cycle of the THESEUS satellite to 80\%~\citep{THESEUS:2017wvz} mainly due to a reduction of observing time owing to the passage through the Southern Atlantic Anomaly, a sky coverage fraction of $0.5$, and a flux limit  $F_{min}= 0.2\ \mbox{photon}\ \mbox{cm}^{-2}\ \mbox{s}^{-1}$ in the $50-300$ keV. This configuration favors a source localization with a precision of $\sim 5$ arcmin, but  only within the central 2 sr of the X-Gamma ray Imaging Spectrometer (XGIS) field of view (FOV)~\citep{THESEUS:2017wvz}.  Thus, we follow~\citet{Belgacem:2019tbw} setting the effective number of the detected sources to $1/3$ of the total number of detected short GRBs, and  analysing two cases: first,  the {\em optimistic} case in which all the events detected by XGIS have a measured redshift; and second, the {\em realistic} case in which the redshift is measured only for sources that are localized   with a precision of $\sim 5$ arcmin.
	In Table~\ref{Tab: joint event }, we report the data for the {\em optimistic} and {\em realistic} cases. 
	We estimate a rate of order $5\div 10$ in the {\em realistic} case, a factor at least $\sim 3$ lower than predicted for the THESEUS mission by~\citet{THESEUS:2017wvz}; while in the  {\em optimistic} case we estimate a rate of order $16\div 31$, a factor at least $\sim 1.6$ to  $\sim 2$ lower than predicted for the THESEUS mission by~\citet{Belgacem:2019tbw}.
	Such differences are rather expected since  we assume a different BNS rate function with respect to~\citet{THESEUS:2017wvz}, and a lower value of the local rate $R_m (z=0)$ with respect to~\citet{Belgacem:2019tbw}. We checked that using the same model of~\citet{THESEUS:2017wvz} and the same value of  $R_m (z=0)$ with respect to~\citet{Belgacem:2019tbw}, we recover their estimated number of events.
	Finally, for the sake of completeness, we illustrate in Figure~\ref{fig:event_distr2} the observed event distribution as a function of the redshift for both the {\em optimistic} and the {\em realistic} cases in the upper and bottom row, respectively. The left, central, and right panels correspond to the SNR thresholds 9, 12, and 15, respectively.
	In each panel, we depict the redshift distribution obtained after one, five and ten years of observations with dark red, red, and light red colors as shown in the legend.
	\begin{table}
		\centering
		\begin{tabular}{c|c|c||c|c||c|c}
			\hline
			\multicolumn{7}{c}{ {\bf Optimistic Case}}\\
			\cline{1-7}
			\multirow{2}{*}{\bf{Years}}&\multicolumn{2}{c||}{\bf SNR = 9}  &      \multicolumn{2}{c||}{ {\bf SNR = 12}} & \multicolumn{2}{c}{\bf SNR = 15}\\
			\cline{2-7}
			&  {\bf \# events} & ${\boldmath z_{max}}$  &  {\bf  \# events} & ${\boldmath z_{max}}$  &  {\bf  \# events} & ${\boldmath z_{max}}$ \\
			\hline
			1 & 31  &2.66  &  25& 1.81  & 16 & 1.36 \\
			5 & 166 &  4.03 & 125& 2.85  & 97 & 2.15 \\
			10 &332&  5.08 & 249 & 2.85  & 204 & 2.41 \\
			\hline
			\hline
			\multicolumn{7}{c}{ {\bf Realistic Case}}\\
			\cline{1-7}
			\multirow{2}{*}{\bf{Years}}&\multicolumn{2}{c||}{\bf SNR = 9}  &      \multicolumn{2}{c||}{ {\bf SNR = 12}} & \multicolumn{2}{c}{\bf SNR = 15}\\
			\cline{2-7}
			&  {\bf \# events} & ${\boldmath z_{max}}$  &  {\bf  \# events} & ${\boldmath z_{max}}$  &  {\bf  \# events} & ${ \boldmath z_{max}}$ \\
			\hline
			1 & 10 &2.66  & 8 & 1.18 & 5 & 1.49 \\
			5 & 55 & 3.00  & 41 & 1.95 & 32 & 2.15 \\
			10 & 110 & 4.13 &83  & 2.60 & 68 & 2.19 \\
			\hline
		\end{tabular}
		\caption{Numbers of GW events with electromagnetic counterpart recorded in each mock catalog depending on the threshold in the SNR and the total number of observational years. We report numbers for both the {\em optimistic} and the {\em realistic} cases.}
		\label{Tab: joint event }
	\end{table}
	
	We checked that changing the model of the luminosity profile of the jet in Eq.~\eqref{jet_profile} does not change remarkably the number of the detected events. We have compared the model in Eq.~\eqref{jet_profile} with a  two-component jet model and a structured jet model in which the energy is a function of the jet angle outside a uniform core~\citep{Lamb2017}. Our results indicates that, using Eq.~\eqref{jet_profile}, 10 and 31 events are detected in the realistic and optimistic cases, respectively. Whereas, using a  two-component jet model one detects 12 and 36 events, respectively. And, finally, using a  structured jet one detects 12 and 37 events, respectively.
	
	\section{Bayesian statistical analysis}\label{sec:four}
	
	To evaluate the accuracy down to which ET will be able to constrain cosmological parameters, we carry out a statistical analysis employing a Monte Carlo Markov Chain (MCMC) algorithm. We will use the \texttt{emcee} Python package~\citep{emcee} to constrain the cosmological parameters by fitting the luminosity distance for all our mock catalogs, {\em i.e.} with and without the detection of an electromagnetic counterpart. Using the Bayes' theorem, we express the posterior distribution, $p(\mathcal{H}|\textbf{d})$, of the cosmological parameters $\mathcal{H} = \lbrace H_0 , \Omega_{k,0}, \Omega_{\Lambda,0}\rbrace$,  in terms of the likelihood function, $p(\textbf{d}|\mathcal{H})$, and the prior distribution $p_0 (\mathcal{H})$~\citep{Padilla2021}:
	\begin{equation}
		p(\mathcal{H}|\textbf{d}) \propto p(\textbf{d}|\mathcal{H})p_0 (\mathcal{H}).
	\end{equation}
	Here, we have labeled the dataset with \textbf{d}$\equiv\lbrace d_i \rbrace_{i=1}^{N}$, where $N$ is the number of observations.
	Furthermore, the uniform prior distributions for all the model parameters are: $H_0 \in \mathcal{U}(35, 85)$, $\Omega_{k,0} \in \mathcal{U}(-1, 1)$ and $\Omega_{\Lambda,0} \in \mathcal{U}(0, 1)$.
	It is worth stressing that in our analysis, we neglect the contribution of source spin to the  amplitude of the signal~\citep{Poisson:1995ef,Baird:2012cu}. Finally, 
	the total likelihood of the GW events is the product of the single event likelihood:
	\begin{equation}
		p(\textbf{d}|\mathcal{H}) = \prod_{i=1}^{N} p(d_i|\mathcal{H}).
	\end{equation}
	
	We will examine three different cases: {\bf (I)} including the redshift information from the electromagnetic signal;  {\bf  (II)} including selection effects due to the cut in SNR and flux with respect to {\bf  (I)}; and, finally, {\bf  (III)} using all the GW events detected by the ET, even those without electromagnetic counterpart, the so called {\em dark sirens}. 
	Specifically:
	\begin{description}
		\item[(I):] the information about the redshift of the source comes from the detection of the electromagnetic counterpart allowing to break the mass-redshift degeneracy. In such a case, we define  and use the following  log-likelihood
		\begin{equation}\label{likelihood:chi2}
			-\log{p(d_i|\mathcal{H})} =  \frac{1}{2}\left(\frac{d_i - d_{L}^{th}(z_i,\mathcal{H})}{\sigma_{d_{i}}}\right)^2\ +\log\left(2\pi\sigma_{d_{i}}\right) \,,
		\end{equation}
		where $d_{L}^{th}(z_i,\mathcal{H})$ is the theoretical luminosity distance given by the Eq.~\eqref{luminosity_distance}, $d_i\equiv d_L(z_i)$, $z_i$ is the redshift of coincident GRB (we assume it perfectly known), and $\sigma_{d_{i}}$ are the mock data and the corresponding error bars, respectively.
		\item[(II):] As in the previous case, the information about the redshift is known.  We use a hierarchical Bayesian framework to include selection effects~\citep{Mandel:2018mve,Vitale:2020aaz}. Generally speaking,
		the posterior probability distribution of a single event in presence of selection effects is~\citep{Mandel:2018mve}
		\begin{equation}
			p(d_i | \mathcal{H})= \frac{\int p(d_i | \lambda)p_{pop}(\lambda | \mathcal{H})d\lambda}{\int p_{det}(\lambda)p_{pop}(\lambda | \mathcal{H})d\lambda},
		\end{equation}
		where $\lambda$ can be whatever parameter related to the GW event, such as spins, masses, and redshift among the others, $p(d_i| \lambda)$ is the likelihood for detecting an event at the distance $d_i$ given the set of parameters $\lambda$, $p_{det}(\lambda| \mathcal{H})$ is the probability of detecting a event with an intrinsic parameter $\lambda $ given the set of parameters $\mathcal{H}$ and, finally, $p_{pop}(\lambda|\mathcal{H})$ is a population modelled prior distribution.
		The denominator is a normalization factor that takes into account the selection effects. In our catalogs, we have two selection effects that may affect final constraints: the first is generated by the selection in the SNR; and the second, by the selection in electromagnetic flux. Therefore, the probability distribution can be written as~\citep{DelPozzo:2011vcw, Mandel:2018mve, Ye:2021klk}
		\begin{equation}
			p(d_i | \mathcal{H})= \frac{\int  p(d_i|D_L)p_{pop}(D_L |z_i , \mathcal{H})d D_L}{\int p_{det}(D_L)p_{pop}(D_L|z_i , \mathcal{H}) dD_L},
		\end{equation} 
		where $p_{pop}(D_L |z_i, \mathcal{H})=\delta(D_L - d_{L}^{th}(z_i,\mathcal{H}))$~\citep{DelPozzo:2011vcw}. The integration of the previous equation leads to 
		\begin{equation}
			p(d_i |\mathcal{H} )= \frac{p(d_i|d_{L}^{th}(z_i,\mathcal{H}))}{p_{det}(d_{L}^{th}(z_i,\mathcal{H}))},
		\end{equation}
		where the numerator is given by Eq.~\eqref{likelihood:chi2}, and the denominator  gives the joint probability of detecting an event and the corresponding short GRB above a certain threshold in SNR and flux:
		\begin{equation}
			p_{det}(d^{th}_L(z_i,\mathcal{H}) | \rho_i > \rho_{t},  F_i > F_{min})=p_{det}^{\rho_i}\,
			p_{det}^{F_i}\,.
		\end{equation}
		The first term gives the probability of detecting an event at redshift $z_i$ and with SNR$=\rho_i$ above a a certain threshold $\rho_t$. From the relation $\rho_i \propto d_{i}^{-1}$, we get
		
		\begin{align}
			p_{det}^{\rho_i}\equiv& p_{det}(d^{th}_L(z_i,\mathcal{H})| \rho_i > \rho_{t})
			=\int_{\rho_{t}}^{+\infty}d\rho_i\ p(\rho_i|d_{L}^{th}(z_i,\mathcal{H})) \nonumber \\
			=&\frac{1}{2} \left[ 1 +  \mathrm{erf}\left( \frac{ d_{L}^{th}(z_i,\mathcal{H})}{\sqrt{2}\sigma_{d_{i}}}\left(  \frac{\rho_{i}}{\rho_t} - 1 \right) \right) \right]\,.
		\end{align}
		The second term gives the probability of detecting an event above  a certain threshold in flux with a given threshold in SNR. Since the luminosity distance is Gaussian distributed and $F_i\propto d_{i}^{-2}$, it follows 
		\begin{align}
			p_{det}^{F_i}\equiv& p_{det}(d^{th}_L(z_i,\mathcal{H}), F_i > F_{min}|\rho_i > \rho_{t}) \nonumber\\
			=& \int^{+\infty}_{F_{min}}dF_i\ p(F_i|d_{L}^{th}(z_i,\mathcal{H})) \nonumber\\
			=&\frac{1}{4} \left[ 1 + \mathrm{sign}\left(\Delta_i\right)\ \mathrm{erf}\left( \frac{\Delta_i}{\sqrt{2}\sigma_{d_{L}}} \right) \right]\,,
		\end{align}
		where, for sake of convenience, we have defined
		\begin{align}
			\Delta_i =    \sqrt{\frac{L_i}{4\pi F_{min}}} - d_{L}^{th}(z_i,\mathcal{H})\,,
		\end{align}
		with $L_i$ the luminosity of short GRB.
		\item[(III):] As already mentioned in the Introduction, 
		the electromagnetic counterpart will be detected only for
		the $\sim 0.08\%$ of the total number of events detected by the ET. Therefore, it is mandatory to investigate which precision can be achieved using all the remaining data although one has to marginalize on the redshift. One way is to marginalize over the redshift of the potential host galaxies obtained by the localization of the GW event~\citep{Schutz1986,MacLeod:2007jd,DelPozzo:2011vcw}. This approach is called statistical host identification technique.
		By applying this methodology to the 47 GW events with detected SNR$> 11$ reported in the third LIGO-Virgo-KAGRA Gravitational Wave Transient Catalog~\citep{LIGO_Catalog_3}, the ~\citet{LIGO_H0:2021} obtain $H_0 = 68_{-6}^{+8}\ \hbox{km} \  \hbox{s}^{-1} \hbox{Mpc}^{-1}$. 
		In addition to the latter method, one can use the three-dimensional cross-correlation technique for GW sources and galaxies~\citep{Mukherjee2021a,Mukherjee2021b}. Both approaches rely on the following key aspects: {\em (a)} the source sky localization error, {\em (b)} the luminosity distance uncertainty, {\em (c)} the overlapping sky area between GW sources and galaxy surveys, {\em (d)} the accurate redshift estimation of galaxies, and {\em (e)} the redshift distribution of galaxies~\citep{Diaz:2022}.
		In particular, increasing the error on the sky localization suppresses the cross-correlation~\citep{Mukherjee2021a}. Hence, the nearby GW sources have better sky localization and provide a more precise estimation of the redshift. On the other hand, the statistical host identification technique depends on whether the host galaxy of the GW source is present in a galaxy catalog~\citep{gray2020,Bera2020}.
		However, even without an electromagnetic counterpart or a complete galaxies catalog or a galaxy surveys with higher galaxy number density, if the redshift distribution of the population of the GW events is known, for instance using population synthesis simulations, one can extract cosmological information~\citep{Ding:2018zrk,Ye:2021klk}. Indeed, the probability of detecting an event at redshift $z_i$ in a specified cosmological model is given by:
		\begin{align}
			p(d_i|\mathcal{H}) &= \int_{0}^{z_{max}} p(d_i , z_i |\mathcal{H}) dz_i \nonumber\\
			\label{dark_likelihood}    &= \int_{0}^{z_{max}} p(d_i|d_L^{th}(z_i,\mathcal{H})) p_{obs}(z_i|\mathcal{H})dz_i.
		\end{align}
		The probability prior distribution of the redshift, $p_{obs}(z_i|\mathcal{H})$, is obtained from the observed events  and already includes detector selection effects~\citep{Ding:2018zrk}, since it is  related to the distribution  $p(z)$ in Eq.~\eqref{red_distr} through the condition that $\rho > \rho_{t}$.
		Furthermore, the likelihood $ p(d_i|d_L^{th}(z_i,\mathcal{H}))$ is  given in Eq.~\eqref{likelihood:chi2}. 
		
		Additionally, it has been recently shown that the redshift of a GW event can be obtained from the measurements of the tidal effects in BNS~\citep{messenger:Read,Ghosh2022}.
		Having such a prior estimation of the redshift directly from the waveform would allow to assume a Gaussian prior distribution on redshift, $p_{obs}(z_i|\mathcal{H})=\mathcal{N}(z_i^{obs}, \sigma_{z_i})$, that would be centered in its corresponding estimated value $z_i^{obs}$ with a variance given by statistical error $\sigma_{z_i}$~\citep{JiangYagi}.
		
		In our analysis, we refer to {\em dark sirens} when using all the events without the electromagnetic counterpart, assuming known the redshift prior distribution from  population synthesis simulations and/or the measure of the tidal deformability parameter. In this case, we adopt the likelihooh defined in Eq.~\eqref{dark_likelihood}.
	\end{description}

	\section{Results}\label{sec:five}
	
	We have built mock catalogs of the luminosity distance extracted by the GW events that will be detected by the forthcoming ET throughout one, five and ten years of observations. Our {\em baseline} model is a flat-$\Lambda$CDM with the fiducial values of parameter $H_0, \Omega_k,$ and $\Omega_\Lambda$ given in Eq.~\eqref{fid_cosmo}. In our mock catalogs, we retained only the events with a SNR above a given threshold, {\em i.e.} SNR$>[9, 12, 15]$. Therefore, we finally have nine mock catalogs with the number of events ranging from  $\sim 10^4$ to $\sim 5\times 10^5$ depending on SNR and on the years of observations, as shown in Table~\ref{Tab:events}. Additionally, we extract from each catalog those events that will have an electromagnetic emission detectable by the THESEUS mission. Therefore, we create other 18 mock catalogs only containing GW events with a detected electromagnetic counterpart. As reported in Table~\ref{Tab: joint event }, to build nine of those catalogs, we assume that all GW events detectable by THESEUS will be effectively detected ({\em optimistic case}), while to build the other nine catalogs we assume that the electromagnetic counterpart is detected in  1/3 of the total number of detectable events ({\em realistic case}). Then, we use a  MCMC algorithm to forecast the precision down to which ET would recover cosmological parameters. We carry out three analyses: {\bf (I)} considering only the events with a detected electromagnetic counterpart, {\bf (II)} the same but including selection effects as explained in the previous section, and {\bf (III)} including {\em dark sirens}.
	We will refer our discussions to the mock catalog that collects   all the events with SNR$>9$ detected throughout ten years of observations (hereby called {\em reference catalog}).
	
	\subsection{Analysis of GW events with a detected electromagnetic counterpart }
	\begin{table*}
		\centering
		\begin{tabular}{c|c|c|c|c||c|c|c}
			\hline
			\multirow{2}{2.5em}{{\bf SNR}}&\multirow{2}{2.5em}{{\bf years}}&\multicolumn{3}{c||}{{\bf Realistic case}} &\multicolumn{3}{c}{{\bf Optimistic case}}\\
			\cline{3-8}
			&& ${\boldmath H_0}$ & ${\boldmath \Omega_{k,0}}$  & ${\boldmath \Omega_{\Lambda,0}}$ &${\boldmath H_0}$ & ${\boldmath \Omega_{k,0}}$  & ${\boldmath \Omega_{\Lambda,0}}$\\
			\hline
			\multirow{3}{2em}{9} & 1 &$66.24_{-1.46}^{+1.44}$ & $-0.04_{-0.30}^{+0.40}$ & $0.67_{-0.35}^{+0.24}$& $66.96_{-1.09}^{+1.05}$ & $0.15_{-0.32}^{+0.34}$ & $0.50_{-0.28}^{+0.25}$\\
			& 5  & $67.51_{-0.83}^{+0.79}$ & $0.10_{-0.21}^{+0.26}$ & $0.62_{-0.19}^{+0.17}$& $67.56_{-0.47}^{+0.45}$ & $-0.02_{-0.11}^{+0.11}$ & $0.70_{-0.09}^{+0.08}$\\
			& 10  & $67.38_{-0.84}^{+0.85}$ & $-0.08_{-0.16}^{+0.17}$ & $0.71_{-0.15}^{+0.13}$& $67.47_{-0.40}^{+0.39}$ & $-0.08_{-0.09}^{+0.08}$ & $0.72_{-0.07}^{+0.07}$\\
			\hline
			\hline
			\multirow{3}{2em}{12} & 1 &$68.05_{-1.10}^{+1.13}$ & $0.11_{-0.30}^{+0.50}$ & $0.71_{-0.33}^{+0.21}$& $66.95_{-1.17}^{+1.32}$ & $0.08_{-0.32}^{+0.41}$ & $0.59_{-0.31}^{+0.25}$\\
			& 5  &$67.45_{-0.84}^{+0.99}$ & $-0.06_{-0.24}^{+0.30}$ & $0.71_{-0.24}^{+0.17}$& $67.63_{-0.36}^{+0.35}$ & $-0.07_{-0.15}^{+0.16}$ & $0.70_{-0.11}^{+0.11}$\\
			& 10 &$67.42_{-0.34}^{+0.36}$ & $0.05_{-0.17}^{+0.19}$ & $0.59_{-0.12}^{+0.12}$& $67.42_{-0.44}^{+0.45}$ & $0.02_{-0.13}^{+0.13}$ & $0.63_{-0.10}^{+0.10}$\\
			\hline
			\hline
			\multirow{3}{2em}{15} & 1  &$67.90_{-1.33}^{+1.20}$ & $-0.09_{-0.36}^{+0.48}$ & $0.70_{-0.34}^{+0.22}$& $67.84_{-0.51}^{+0.51}$ & $-0.08_{-0.34}^{+0.47}$ & $0.69_{-0.29}^{+0.22}$\\
			& 5 &$67.33_{-0.92}^{+0.93}$ & $0.26_{-0.34}^{+0.36}$ & $0.49_{-0.24}^{+0.24}$& $68.13_{-0.66}^{+0.61}$ & $-0.01_{-0.19}^{+0.23}$ & $0.68_{-0.15}^{+0.14}$\\
			& 10  &$68.10_{-0.72}^{+0.68}$ & $-0.06_{-0.20}^{+0.25}$ & $0.74_{-0.18}^{+0.14}$& $67.63_{-0.62}^{+0.57}$ & $0.05_{-0.17}^{+0.18}$ & $0.62_{-0.13}^{+0.13}$\\
			\hline
		\end{tabular}
		\caption{The median value and the 68\% confidence level of the posterior distributions of the  parameters of our {\em baseline} model for each SNR and for one, five and ten years of observations, as obtained from the MCMC analyses carried out on mock catalog collecting  the GW events with a detected electromagnetic counterpart.}
		\label{Tab:CASE I}
	\end{table*}
	
\begin{figure*}
		\centering
		\includegraphics[width=0.48\textwidth]{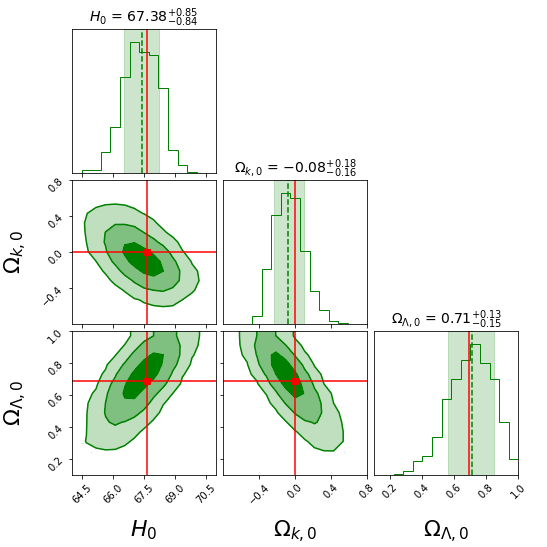}
		\includegraphics[width=0.48\textwidth]{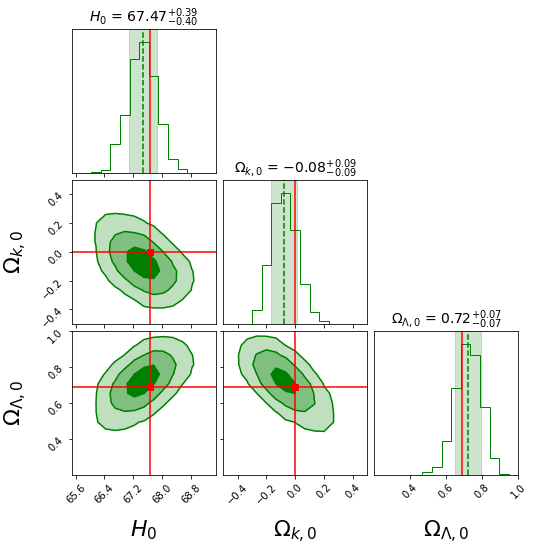}
		\caption{ The figure illustrates the 68\%, 95\% and 99.7\% of confidence level obtained from the posterior distribution of the parameters of our {\em baseline} model. The left and right panels show the results obtained from the {\em realistic} and {\em optimistic} analyses, respectively, carried out on the events detected after ten years of observations and using all the events with SNR$>9$. The vertical red line in the histograms and a red point in the contour plot indicate the true values of the corresponding cosmological parameter. While, the vertical dashed line indicates the median value and the shaded band indicates the $1\sigma$ confidence interval. }
		\label{fig:MCMC_results_CASE_I}
	\end{figure*}

	Table~\ref{Tab:CASE I} reports the median value and the 68\% confidence level of the posterior distributions of the  parameters of our {\em baseline} model for each SNR and for one, five and ten years of observations. Those results are obtained by applying our pipeline to the mock catalog of GW events with a detected electromagnetic counterpart in both {\em realistic} and {\em optimistic} cases. It is worth remembering that the number of events we predicted to be detected ranges from a few events at redshift around two  to at most $\sim 110$ events at a maximum redshift of $\sim 4$ in the  {\em realistic} case. While, in the  {\em optimistic} case, we found  a number of events ranging from $\sim 30$ events  to  $\sim 330$ events at a maximum  redshift of $\sim 5$ (as shown in Table~\ref{Tab: joint event }). Therefore, final results will be affected by the limited number of events and the relatively small redshift at which sometimes they are located. Said that, we predicted that the Hubble constant may be detected with an accuracy at most of $\sigma_{H_0}\approx 0.85$ and $0.40$ km s$^{-1}$ Mpc$^{-1}$ for our {\em reference catalog} representing an uncertainty at level of $\sim 0.7\%$ and $\sim 0.4\%$
	in the {\em realistic} and {\em optimistic} cases, respectively.  On the other hand, in the same analysis, the curvature parameter $\Omega_{k,0}$ may be detected with an accuracy at most of $\sigma_{\Omega_{k,0}}\approx 0.17$ or $0.09$, while the cosmological constant density with an accuracy at most of $\sigma_{\Omega_{\Lambda,0}}\approx 0.14$ or $0.07$ in the {\em realistic} and {\em optimistic} cases, respectively. We show in Figure~\ref{fig:MCMC_results_CASE_I} the contour plots corresponding to the  68\%, 95\% and 99.7\% of confidence level obtained from the posterior distributions of the parameters of our {\em baseline} model. The left and right panels show the analysis in the {\em realistic} and {\em optimistic} case, respectively, carried out on the {\em reference catalog}. We depict the input value of the cosmological parameters as a vertical red line in the histograms and a red point in the contour plot. While, in the histograms, the vertical dashed line indicates the median value and the shaded band indicates the $1\sigma$ confidence interval. The true values of the cosmological parameters are always recovered within $1\sigma$. 
	For sake of completeness we report all the other contour plots in the Appendix~A of Supplementary Materials (SM).

	\begin{table*}
		\centering
		\begin{tabular}{c|c|c|c|c||c|c|c}
			\hline
			\multirow{2}{2.5em}{{\bf SNR}}&\multirow{2}{2.5em}{{\bf years}}&\multicolumn{3}{c||}{{\bf Realistic case}} &\multicolumn{3}{c}{{\bf Optimistic case}}\\
			\cline{3-8}
			&&${\boldmath H_0}$ & ${\boldmath \Omega_{k,0}}$  & ${\boldmath \Omega_{\Lambda,0}}$ &${\boldmath H_0}$ & ${\boldmath \Omega_{k,0}}$  & ${\boldmath \Omega_{\Lambda,0}}$\\
			\hline
			\multirow{3}{2em}{9} & 1  &$66.16_{-1.56}^{+1.38}$ & $-0.07_{-0.31}^{+0.47}$ & $0.66_{-0.39}^{+0.25}$& $66.92_{-1.08}^{+1.03}$ & $0.16_{-0.35}^{+0.41}$ & $0.48_{-0.28}^{+0.29}$\\
			& 5  &$67.59_{-0.85}^{+0.83}$ & $0.04_{-0.20}^{+0.25}$ & $0.66_{-0.19}^{+0.16}$& $67.59_{-0.46}^{+0.44}$ & $-0.04_{-0.12}^{+0.12}$ & $0.71_{-0.09}^{+0.08}$\\
			& 10 & $67.49_{-0.87}^{+0.70}$ & $-0.11_{-0.15}^{+0.16}$ & $0.74_{-0.15}^{+0.12}$& $67.46_{-0.39}^{+0.41}$ & $-0.08_{-0.09}^{+0.09}$ & $0.72_{-0.08}^{+0.07}$ \\
			\hline
			\hline
			\multirow{3}{2em}{12} & 1 &$67.94_{-1.16}^{+1.03}$ & $0.12_{-0.33}^{+0.51}$ & $0.71_{-0.34}^{+0.22}$& $67.09_{-1.36}^{+1.18}$ & $0.11_{-0.33}^{+0.35}$ & $0.60_{-0.29}^{+0.25}$\\
			& 5 &$67.61_{-0.93}^{+0.87}$ & $-0.09_{-0.23}^{+0.35}$ & $0.73_{-0.26}^{+0.16}$& $67.57_{-0.37}^{+0.41}$ & $-0.06_{-0.15}^{+0.16}$ & $0.68_{-0.11}^{+0.15}$\\
			& 10 &$67.46_{-0.34}^{+0.35}$ & $0.03_{-0.15}^{+0.18}$ & $0.61_{-0.12}^{+0.10}$& $67.48_{-0.47}^{+0.40}$ & $0.01_{-0.13}^{+0.13}$ & $0.64_{-0.10}^{+0.09}$\\
			\hline
			\hline
			\multirow{3}{2em}{15} & 1 &$67.91_{-1.30}^{+1.08}$ & $-0.11_{-0.32}^{+0.50}$ & $0.73_{-0.36}^{+0.20}$& $67.97_{-0.54}^{+0.42}$ & $-0.15_{-0.29}^{+0.44}$ & $0.73_{-0.28}^{+0.19}$\\
			& 5 & $67.26_{-0.87}^{+0.91}$ & $0.28_{-0.36}^{+0.35}$ & $0.49_{-0.23}^{+0.25}$& $67.97_{-0.64}^{+0.65}$ & $0.06_{-0.22}^{+0.22}$ & $0.65_{-0.15}^{+0.15}$\\
			& 10 &$68.09_{-0.70}^{+0.68}$ & $-0.08_{-0.21}^{+0.22}$ & $0.74_{-0.16}^{+0.15}$& $67.75_{-0.56}^{+0.55}$ & $0.01_{-0.17}^{+0.18}$ & $0.65_{-0.12}^{+0.12}$\\
			\hline
		\end{tabular}
		\caption{The same as Table~\ref{Tab:CASE I} but including the selection effects as discussed in Section~\ref{sec:four}.}
		\label{Tab:CASE II}
	\end{table*}
	
\begin{figure*}
		\centering
		\includegraphics[width=0.49\textwidth]{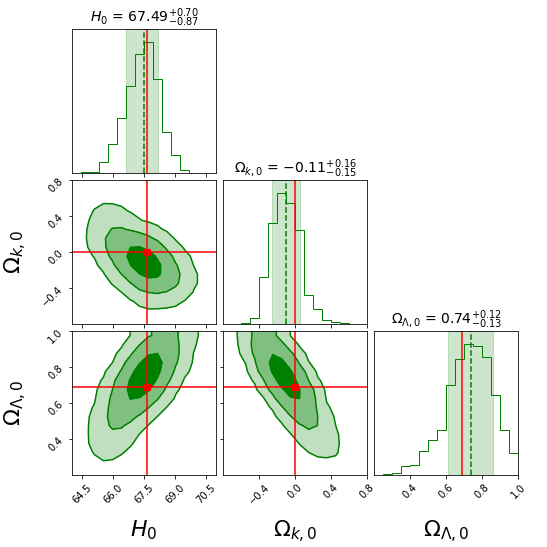}
		\includegraphics[width=0.49\textwidth]{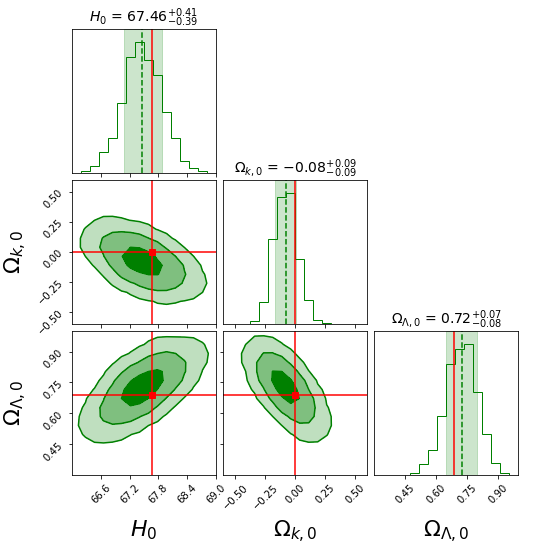}
		\caption{The same as Figure~\ref{fig:MCMC_results_CASE_I} but including the selection effects as discussed in Section~\ref{sec:four}.}
		\label{fig:MCMC_results_CASE_II}
	\end{figure*}

	In addition to the previous results, we repeat all the analyses including the selection effects as discussed in Section~\ref{sec:four}. We note that including selection effects has a very small impact on the final results of the order of a few percent.
	In Figure~\ref{fig:MCMC_results_CASE_II} we show the contour plots corresponding to the  68\%, 95\% and 99.7\% of confidence level obtained from the posterior distributions of the parameters of our {\em baseline} model.
	Panels, lines and colors follow the same design of Figure~\ref{fig:MCMC_results_CASE_I}, while  the true values of the cosmological parameters are again recovered within $1\sigma$. For sake of completeness, we report all the other contour plots in the Appendix~B of SM.

	We can compare our results with the analysis of non-flat $\Lambda$CDM cosmology, carried out by~\cite{DiValentino2021b, Akarsu2021}, based on the CMB power spectrum, Cosmic Chronometers (CC) and Type Ia Supernovae (SNIa).
	\citet{DiValentino2021b} use the CMB power spectrum together with the Baryonic Acoustic Oscillation (BAO) data and the Pantheon sample of SNIa data. When using CMB+BAO data, they constrained the Hubble constant with an accuracy on $\sigma_{H_0}\approx 1.7$  km s$^{-1}$ Mpc$^{-1}$ and the curvature parameter with an accuracy on $\sigma_{\Omega_{k,0}}\approx 0.005$, which degrade to $\sigma_{H_0}\approx 4.4$  km s$^{-1}$ Mpc$^{-1}$ and $\sigma_{\Omega_{k,0}}\approx 0.02$ when using CMB+SNIa. On the other hand, the accuracy on Hubble constant, from~\citet{Akarsu2021}, is $\sigma_{H_0}\approx 3$  km s$^{-1}$ Mpc$^{-1}$ and $\sigma_{H_0}\approx 10$  km s$^{-1}$ Mpc$^{-1}$. While the accuracy on the curvature parameter is $\sigma_{\Omega_{k,0}}\approx 0.16$ and $\sigma_{\Omega_{k,0}}\approx 0.14$, using the CC and Pantheon sample of SNIa, respectively. Our results from the {\em reference catalog} show an improvement of $\sim 40\%$ and $\sim 76\%$ in the accuracy on the Hubble constant, in the {\em realistic} and {\em optimistic} cases, respectively, with respect to the best scenario of~\citet{DiValentino2021b}. While, with respect to~\citet{Akarsu2021}, we predict to improve the accuracy on the Hubble constant of a factor $\sim 3.5$ and $\sim 7.5$ in the {\em realistic} and {\em optimistic} cases, respectively. Moreover, we are not competitive on the accuracy predicted on the curvature parameter by~\citet{DiValentino2021b}. However, for the {\em realistic} case, we predict an accuracy of the same order of~\citet{Akarsu2021}, whereas for {\em optimistic} case we calculate an improvement of a factor $\sim 1.6$.

	\subsection{Analysis of the dark sirens }
	\begin{table}
		\centering
		\begin{tabular}{c|c|c|c|c}
			\hline
			\multicolumn{5}{c}{\textbf{Dark Sirens}} \\
			\hline
			
			\textbf{SNR} & \textbf{years}  & ${\boldmath H_0}$ & ${\boldmath \Omega_{k,0}}$  & ${\boldmath \Omega_{\Lambda,0}}$\\
			\hline
			\multirow{3}{2em}{9} & 1  &$67.70_{-0.12}^{+0.13}$ & $0.00_{-0.02}^{+0.02}$  & $0.67_{-0.02}^{+0.02}$\\
			& 5  &$67.63_{-0.05}^{+0.05}$ & $0.01_{-0.01}^{+0.01}$  & $0.69_{-0.01}^{+0.01}$  \\
			& 10 &$67.68_{-0.04}^{+0.04}$ & $0.00_{-0.01}^{+0.01}$  & $0.69_{-0.01}^{+0.01}$ \\
			\hline
			\hline
			\multirow{3}{2em}{12} & 1 & $67.56_{-0.38}^{+0.21}$ & $-0.02_{-0.03}^{+0.04}$  & $0.70_{-0.03}^{+0.03}$ \\
			& 5  &$67.62_{-0.06}^{+0.06}$ & $0.01_{-0.02}^{+0.01}$  & $0.74_{-0.07}^{+0.08}$\\
			& 10  &$67.68_{-0.05}^{+0.06}$ & $0.00_{-0.01}^{+0.01}$  & $0.69_{-0.01}^{+0.01}$\\
			\hline
			\hline
			\multirow{3}{2em}{15} & 1 &$67.79_{-0.41}^{+0.28}$ & $-0.07_{-0.07}^{+0.07}$  & $0.70_{-0.03}^{+0.03}$ \\
			& 5    & $67.69_{-0.07}^{+0.07}$ & $-0.00_{-0.03}^{+0.03}$  & $0.68_{-0.02}^{+0.02}$\\
			& 10  &$67.67_{-0.06}^{+0.05}$ & $0.01_{-0.01}^{+0.01}$  & $0.69_{-0.01}^{+0.01}$\\
			\hline
		\end{tabular}
		\caption{The same as Table~\ref{Tab:CASE I} but for the dark sirens assuming known the redshift prior distribution.}
		\label{Tab:CASE III}
	\end{table}

	In Table~\ref{Tab:CASE III}, we report the median value and the 68\% confidence level of the posterior distributions of the  parameters of our {\em baseline} model for each SNR and for one, five and ten years of observations. These results are obtained by applying our pipeline to the \emph{dark sirens}. As reported in Table~\ref{Tab:events}, the number of events ranges from $\sim 1\times10^4$ to $\sim 5\times10^5$ and are located at a maximum redshift of $\sim 6.9$ depending on the SNR and the total year of observations. 
	
\begin{figure}
		\centering
		\includegraphics[width=0.49\textwidth]{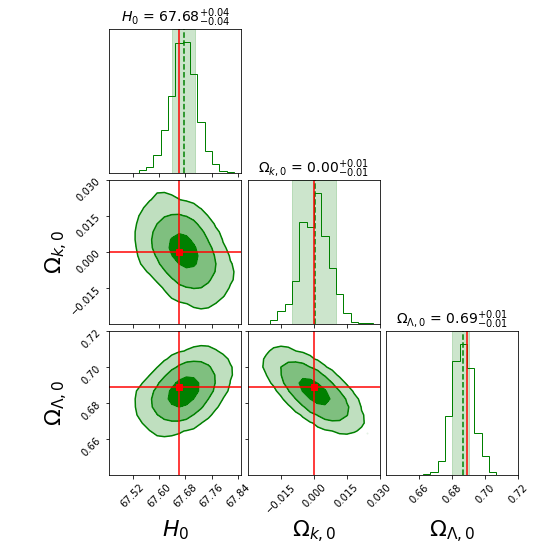}
		\caption{The same of Figure~\ref{fig:MCMC_results_CASE_I} but for the dark sirens assuming known the redshift prior distribution.}
		\label{fig:MCMC_results_CASE_III}
	\end{figure}

	Since the redshift of the event is {\em a priori} unknown, we have to marginalize over it. Thus, adopting the probability distribution given in  Eq.~\eqref{red_distr}, we found that the Hubble constant may be detected with an accuracy of (at most of) $\sigma_{H_0}\approx 0.04$  km s$^{-1}$ Mpc$^{-1}$ using our {\em reference catalog}. This would represent an uncertainty at level of $\sim 0.06\%$, about a factor 10 better than using only events with an electromagnetic counterpart. On the other hand, in the same analyses, the curvature parameter $\Omega_{k,0}$ may be detected with an accuracy of $\sigma_{\Omega_{k,0}}\approx 0.01$. While, the cosmological constant density with an accuracy  of $\sigma_{\Omega_{\Lambda,0}}\approx 0.01$, which represent an improvement of about a factor $\sim 8$.  As a comparison Planck Collaboration constrains the Hubble constant with an accuracy of $\sigma_{H_0}\approx 0.5$  km s$^{-1}$ Mpc$^{-1}$, and the cosmological constant density with an accuracy of $\sigma_{\Omega_{\Lambda,0}}\approx 0.007$. While, the curvature parameter is constrained with an accuracy of $\sigma_{\Omega_{k,0}}\approx 0.02$~\citep{Aghanim2020}. In Figure~\ref{fig:MCMC_results_CASE_III}, we show the contour plots corresponding to the  68\%, 95\% and 99.7\% of confidence level obtained from the posterior distributions of the parameters of our {\em baseline} model.
	Lines and colors follow the same design of Figure~\ref{fig:MCMC_results_CASE_I}, and  the true values of the cosmological parameters are again recovered within $1\sigma$ but with an increased accuracy.
	For the sake of completeness, we report all the other contour plots in the Appendix~C of SM. 
	
	The huge improvements show the full potential of the dark sirens. We can compare the results of Table~\ref{Tab:CASE III} with those of~\citet{DelPozzo:Darksirens}. On the cosmological densities, we confirm the order of magnitudes of the accuracy extrapolated by~\citet{DelPozzo:Darksirens}. While on the Hubble constant, we note that the accuracy predicted by~\citet{DelPozzo:Darksirens} is much better than in our results analysis. One possible reason is that the SFR is assumed to be a constant giving more events at low redshift where the model is more sensitive to the value of the Hubble constant.
	
	Our results for the {\em reference catalog} show an improvements of a factor of $\sim 75$ and $\sim 42$ in the accuracy on the Hubble constant with respect to the best constraints of~\citet{Akarsu2021} and~\citet{DiValentino2021b}, respectively. On the other hand, we improve a factor of  $\sim 16$ in the accuracy on the curvature parameter with respect to the best constraint of~\citet{Akarsu2021}, while we are comparable with the analysis of~\citet{DiValentino2021b}.
	
	Additionally, we also repeat the analysis adopting, to marginalize over the redshift, a  Gaussian prior distribution centered on the {\em fiducial} redshift as it was obtained from the measurements of the tidal effects in BNS~\citep{messenger:Read,JiangYagi}. As expected, even having partial information on the redshift with a statistical error of about $10\%$, which depends on neutron stars equation of state, the final results improve a factor 1.4 in the Hubble constant, $\sigma_{H_0}\approx 0.03$  km s$^{-1}$ Mpc$^{-1}$, and a factor 2.5 in the curvature parameter, $\sigma_{\Omega_{k,0}}\approx 0.004$, and in the cosmological constant density, $\sigma_{\Omega_{\Lambda,0}}\approx 0.004$.

	\subsection{Impact of other star formation rates and the time delay distribution }\label{sec:check}
	
	Our {\em baseline} model adopts the {\em Vangioni model} for the SFR and the  {\em power law} form of the time delay distribution. Here, we show that making other choices leads to similar constraints on the cosmological parameters. We consider three different models:  {\bf Model 1} based on the {\em Vangioni model} for the SFR and the  {\em exponential distribution}  of the time delay distribution; {\bf Model 2} based on the {\em Madau - Dickison model} for the SFR and the  {\em power law} form of the time delay distribution; {\bf Model 3} based on the {\em Madau - Dickison model} for the SFR and the  {\em exponential distribution}  of the time delay distribution.  For each model we repeat the analysis with SNR$>9$ and an observational period of ten years for both dark sirens and events with electromagnetic counterpart. 
	\begin{table}
		\centering
		\begin{tabular}{c|c|c|c|c}
			\hline
			\multicolumn{5}{c}{{\bf GW + EM events}}\\
			\hline
			{\bf MODEL}& \textbf{\# events} &${\boldmath H_0}$ & ${\boldmath \Omega_{k,0}}$  & ${\boldmath \Omega_{\Lambda,0}}$ \\
			\hline
			\textbf{Baseline model} & 332 &$67.47_{-0.40}^{+0.39}$ & $-0.08_{-0.09}^{+0.08}$ & $0.72_{-0.07}^{+0.07}$\\
			
			\textbf{Model 1} & 603 &$67.18_{-0.32}^{+0.34}$ & $0.01_{-0.07}^{+0.07}$ & $0.65_{-0.06}^{+0.06}$\\
			
			\textbf{Model 2} & 271 &$67.48_{-0.30}^{+0.30}$ & $-0.09_{-0.10}^{+0.09}$ & $0.71_{-0.07}^{+0.07}$\\
			
			\textbf{Model 3} & 536 &$67.20_{-0.28}^{+0.27}$ & $0.01_{-0.07}^{+0.08}$ & $0.65_{-0.06}^{+0.05}$\\
			\hline
			\hline
			\multicolumn{5}{c}{{\bf Dark Sirens}}\\
			\hline
			{\bf MODEL}& \textbf{\# events} &${\boldmath H_0}$ & ${\boldmath \Omega_{k,0}}$  & ${\boldmath \Omega_{\Lambda,0}}$ \\
			\hline
			\textbf{Baseline model} & 521552 &$67.68_{-0.04}^{+0.04}$ & $0.00_{-0.01}^{+0.01}$  & $0.69_{-0.01}^{+0.01}$\\
			
			\textbf{Model 1} &1143212 &$67.64_{-0.04}^{+0.04}$ & $0.00_{-0.01}^{+0.01}$  & $0.69_{-0.01}^{+0.01}$\\
			
			\textbf{Model 2} & 443560 & $67.62_{-0.05}^{+0.05}$ & $0.01_{-0.01}^{+0.01}$ & $0.68_{-0.01}^{+0.01}$ \\
			\textbf{Model 3} & 966659&$67.68_{-0.04}^{+0.04}$ & $-0.01_{-0.01}^{+0.01}$  & $0.68_{-0.01}^{+0.01}$\\
			\hline
		\end{tabular}
		\caption{The {\em baseline} model adopts the {\em Vangioni model} for the SFR and the  {\em power law} form of the time delay distribution; {\bf Model 1} is based on the {\em Vangioni model} for the SFR and the  {\em exponential distribution}  of the time delay distribution; {\bf Model 2} is based on the {\em Madau - Dickison model} for the SFR and the  {\em power law} form of the time delay distribution; {\bf Model 3} is based on the {\em Madau - Dickison model} for the SFR and the  {\em exponential distribution}  of the time delay distribution. }
		\label{Tab:check}
	\end{table}
	
	In Table~\ref{Tab:check}, we report for each model the median values and the corresponding 68\% of confidence level of the cosmological parameters. For both analyses the one considering only GW events with electromagnetic counterpart, and the other one using {\em dark sirens}. We found that the relative errors on the cosmological parameters are not affected by the particular choice of neither the SFR and the time delay distribution. 
	
	\section{Discussion and Conclusions}\label{sec:six}
	
	We used mock catalogs of GW events from BNSs to forecast the precision down to which ET will be able to constrain the Hubble constant $H_0$ and the curvature parameter $\Omega_{k,0}$. Although there were other analyses carried out to measure the accuracy on the Hubble constant achievable with ET~\citep{Zhao:2010sz,Cai:2017,Belgacem:2019tbw,Ssohrab:2020,Yang:2021qge,Herbett:2022}, these focused on a flat-$\Lambda$CDM model. Therefore, the forecasts for a non-flat $\Lambda$CDM cosmology  are currently missing. Hence, we built nine mock catalogs containing GW events for one, five, and ten years of observational runs, and for three different thresholds in SNR. Additionally, starting from each one of those nine mock catalogs, we extracted a mock catalog of GW events with associated (and detected) electromagnetic counterpart using THESEUS satellite. In Table~\ref{Tab:events} and~\ref{Tab: joint event }, we report the number of events expected for each combination of SNR and observational run.

	This is a proof-of-principle study and the accuracy reached in our analysis depends on the assumptions made to build the mock catalogs. Therefore, we report a list of such assumptions for sake of clearness:\begin{enumerate}
		\item we set the SFR to Eq.~\eqref{VAN:SFR}, and the time delay distribution to $P(t_d) \propto t_{d}^{-1}$. The model parameters of both are fixed;
		\item we set the value of $R_{m}(z=0)$ to the one obtained by the LIGO/Virgo/Kagra collabaration~\cite{LIGOScientific:2021psn}, and consequently, we assume that the NS masses range uniformly in the interval [1,2.5] $M_{\odot}$;
		\item we use the latest power spectral density for ET;
		\item we assume that the only contributions to luminosity distance error are given by~\eqref{sigma_dl};
		\item in the modelling of the GW likelihood, we neglect of effect of GW detector sensitivity on the sky-position of the sources;
		\item we assume a Gaussian profile of the GRB jet, Eq.~\eqref{jet_profile} and the distribution of GRB peak luminosity, Eq.~\eqref{distr_Lp};
		\item the number of combined events is strictly related to the features of THESEUS satellite.
	\end{enumerate}
	
	Once mock catalogs have been compiled, we run our MCMC pipeline to forecast the accuracy down to which ET can recover the fiducial cosmological model used to build the mock catalog themselves. 
	We run two sets of analyses: one that focused on catalogs of GW events with a detected electromagnetic counterpart, and another that focused on the whole mock catalogs (including also events without a detected electromagnetic counterpart). In our best scenario, using the mock catalog that collects all the events detected after ten years of observations and using all the events with SNR$>9$, we obtain that ET can constrain the cosmological parameters with an accuracy of $\sigma_{H_0}\approx 0.40$ km s$^{-1}$ Mpc$^{-1}$, $\sigma_{\Omega_{k,0}}\approx 0.09$, and  $\sigma_{\Omega_{\Lambda,0}}\approx 0.07$, using only events with a detected electromagnetic counterpart. Our results show that ET may improve of $\sim 76\%$ the accuracy on the Hubble constant with respect to the best scenario of~\citet{DiValentino2021b}, and a factor of $\sim 7.5$ with respect to~\citet{Akarsu2021}. 
	When we consider all the GW events without the electromagnetic counterpart, {\em i.e. dark sirens}, we found that  ET may constrain the cosmological parameters with an accuracy of $\sigma_{H_0}\approx 0.04$ km s$^{-1}$ Mpc$^{-1}$, $\sigma_{\Omega_{k,0}}\approx 0.01$, and  $\sigma_{\Omega_{\Lambda,0}}\approx 0.01$. The improvements with respect to the previous analyses is due to the larger number of events under consideration, which mitigate not knowing the redshift {\em a priori}. Finally, we found that ET would improve of factor $\sim 75$ and $\sim 42$  the accuracy on the Hubble constant with respect to the best constraints of~\citet{Akarsu2021} and~\citet{DiValentino2021b}, respectively, and a factor of $\sim 16$ in the accuracy on the curvature parameter with respect to the best constraint of~\citet{Akarsu2021}. These results show the huge potential of ET to strongly improve current constraints on the cosmological parameters of non-flat $\Lambda$CDM cosmology as 
	they will be affected by different systematics compared to the analyses based on classical electromagnetic  standard candles.

	\section*{Acknowledgments}
	
	MC, DV, and SC acknowledge the support of Istituto Nazionale di Fisica Nucleare (INFN) iniziative specifiche MOONLIGHT2, QGSKY, and TEONGRAV.
	IDM acknowledges support from Ayuda  IJCI2018-036198-I  funded by  MCIN/AEI/  10.13039/501100011033  and:  FSE  “El FSE  invierte  en  tu  futuro”  o  “Financiado  por  la  Unión  Europea   “NextGenerationEU”/PRTR. 
	IDM is also supported by the projects PGC2018-096038-B-I00 and PID2021-122938NB-I00 funded by the Spanish "Ministerio de Ciencia e Innovación" and FEDER “A way of making Europe", and by the project SA096P20 Junta de Castilla y León.
	DV also acknowledges the FCT project with ref. number
	PTDC/FIS-AST/0054/2021.
	
	\subsection*{Data Availability Statement}
	No new data were generated or analysed in support of this research.
	
	\bibliographystyle{mnras}
	\bibliography{Biblio} 
	
	\newpage
	\onecolumn

\appendix
\section{Contours plots relative to the analysis of GW events with a detected electromagnetic counterpart }\label{App:A}

\begin{figure}[H]
    \centering
    \includegraphics[width=0.33\textwidth]{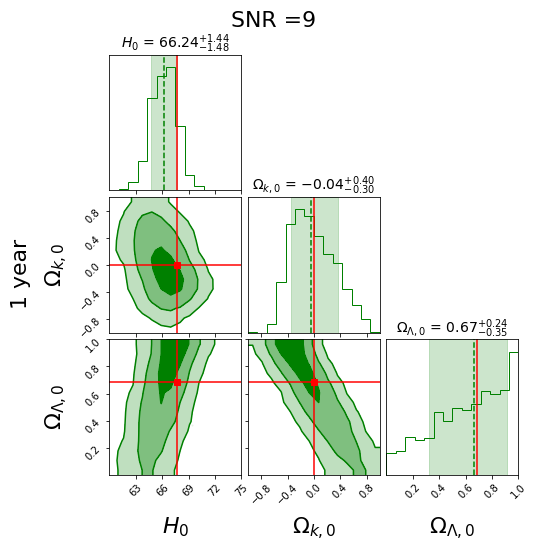}
    \includegraphics[width=0.33\textwidth]{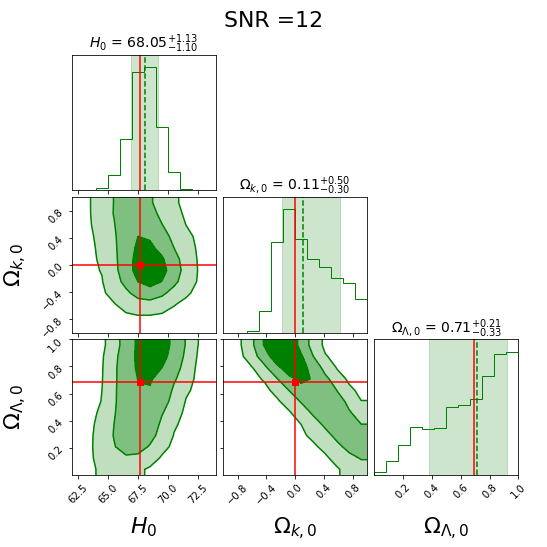}
    \includegraphics[width=0.32\textwidth]{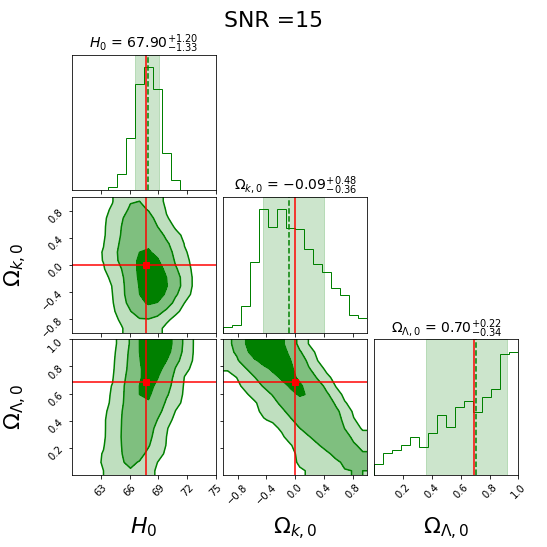}\\
    \includegraphics[width=0.33\textwidth]{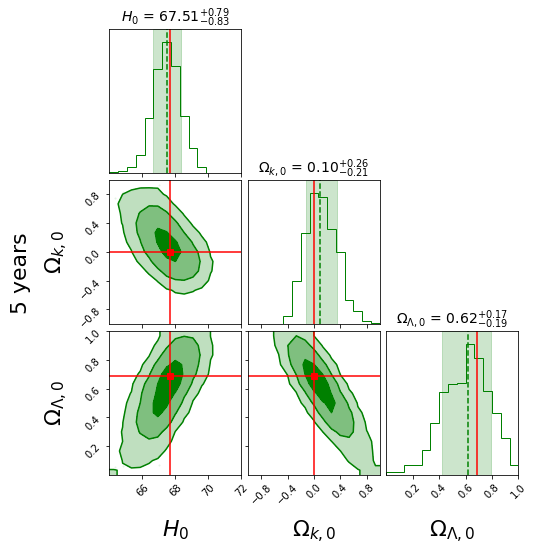}
    \includegraphics[width=0.33\textwidth]{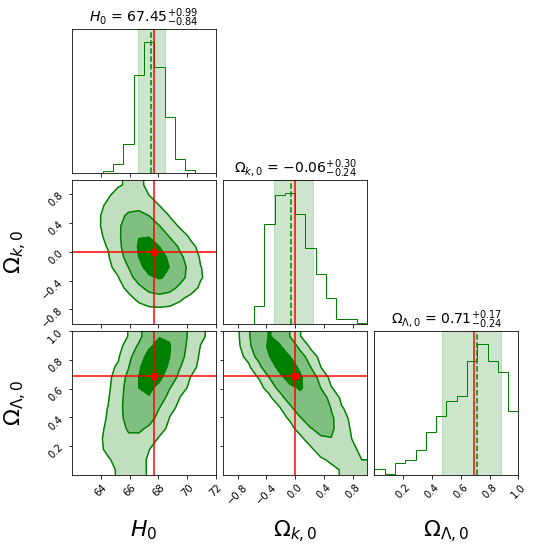}
    \includegraphics[width=0.32\textwidth]{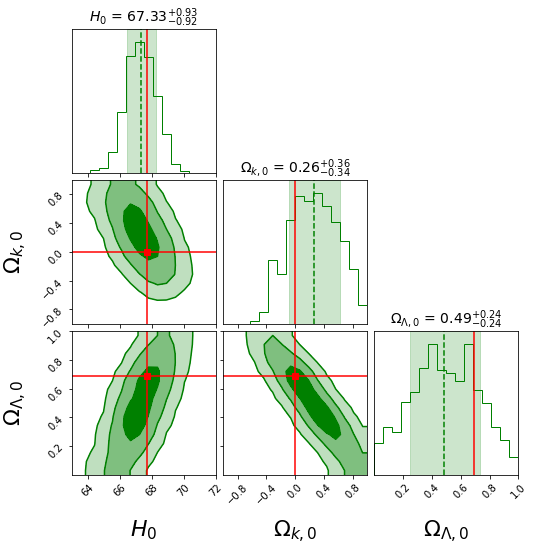}\\
    \includegraphics[width=0.33\textwidth]{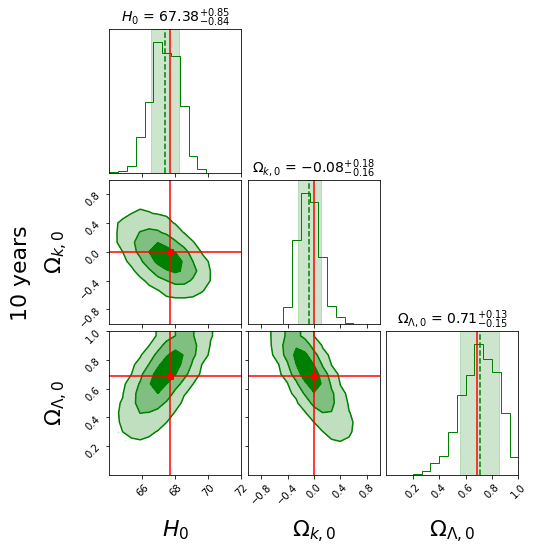}
    \includegraphics[width=0.33\textwidth]{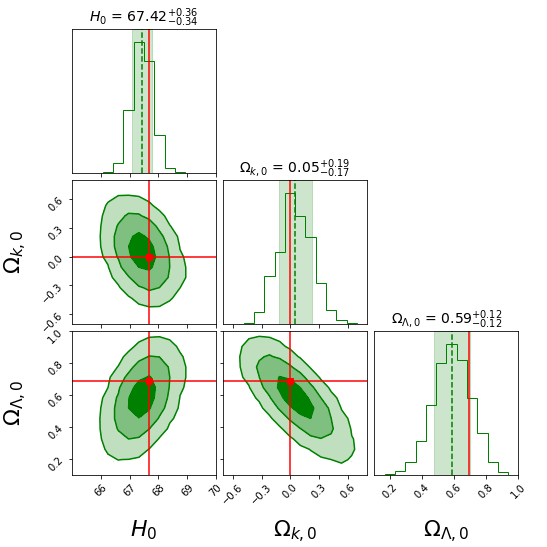}
    \includegraphics[width=0.32\textwidth]{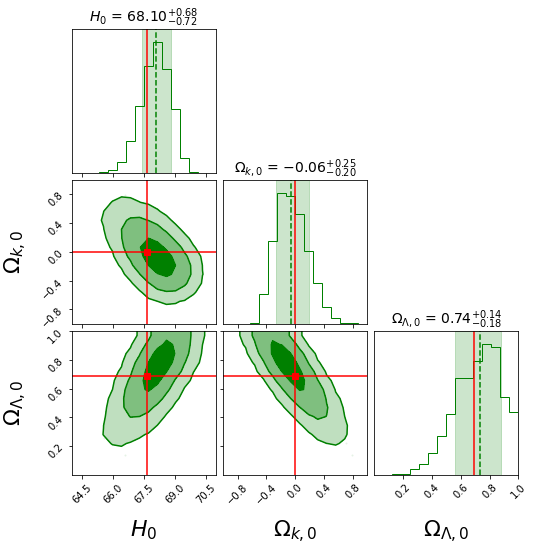}
    \caption{The figure illustrates the 68\%, 95\% and 99.7\% of confidence level obtained from the posterior distribution of the parameters of our {\em baseline} model. The panels show the results obtained from the {\em realistic} analysis carried out on each catalog constructed. The vertical red line in the histograms and a red point in the contour plot indicates the true values of the corresponding cosmological parameter. While, the vertical dashed line indicates the median value and the shaded band indicates the $1\sigma$ confidence interval. }
    \label{fig:contours_realistic_CASE_I}
\end{figure}

\begin{figure}
    \centering
    \includegraphics[width=0.33\textwidth]{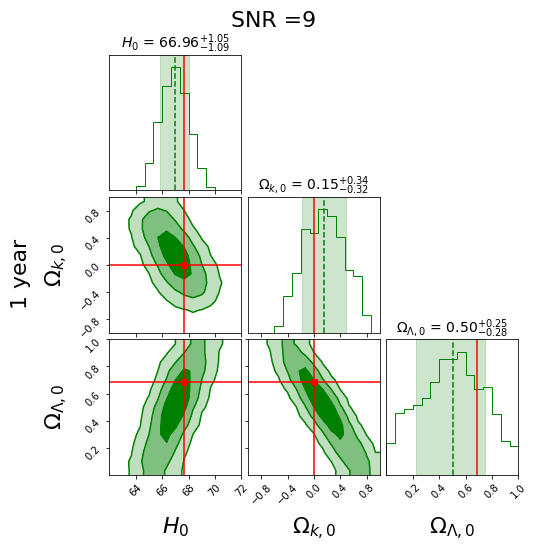}
    \includegraphics[width=0.33\textwidth]{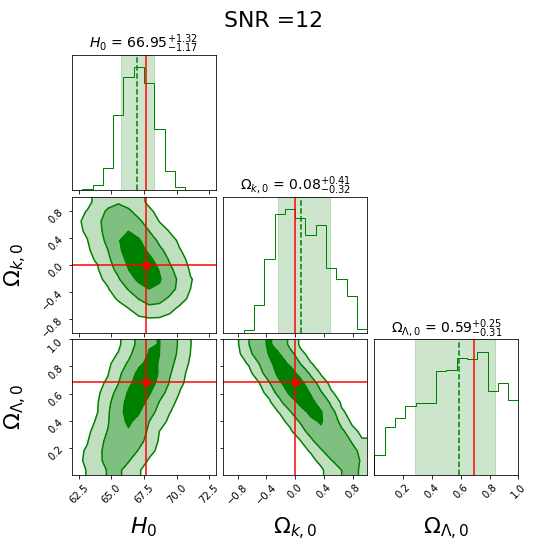}
    \includegraphics[width=0.32\textwidth]{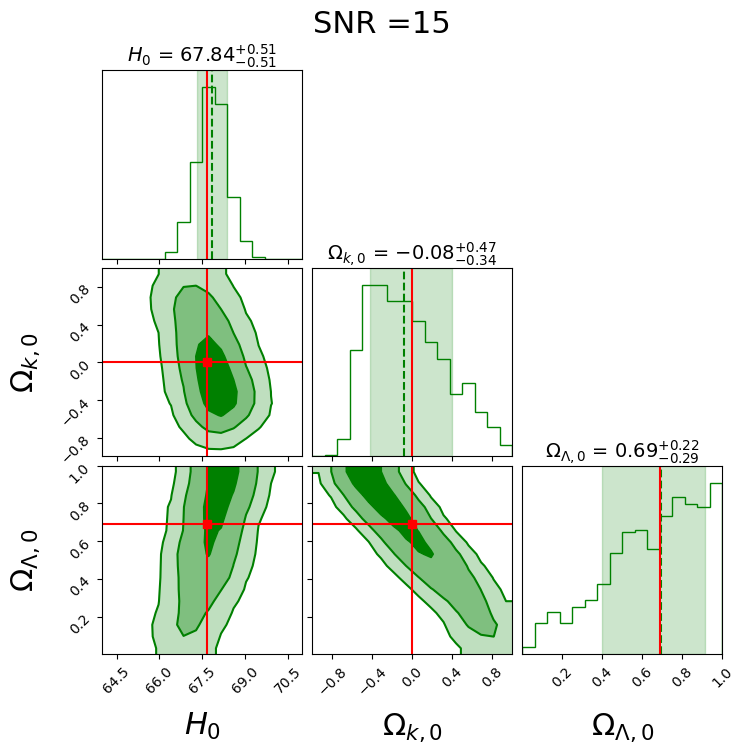}\\
    \includegraphics[width=0.33\textwidth]{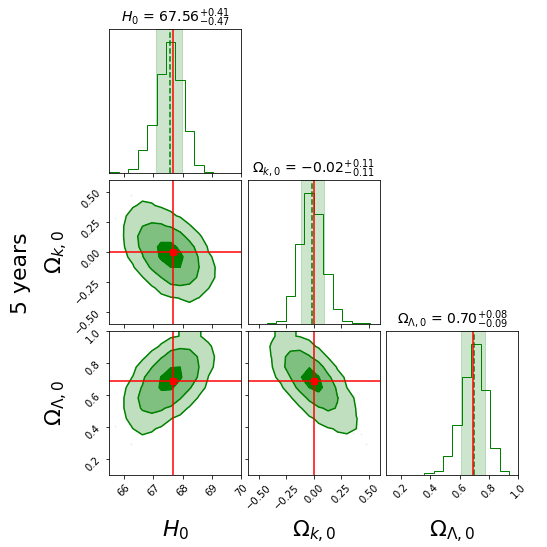}
    \includegraphics[width=0.33\textwidth]{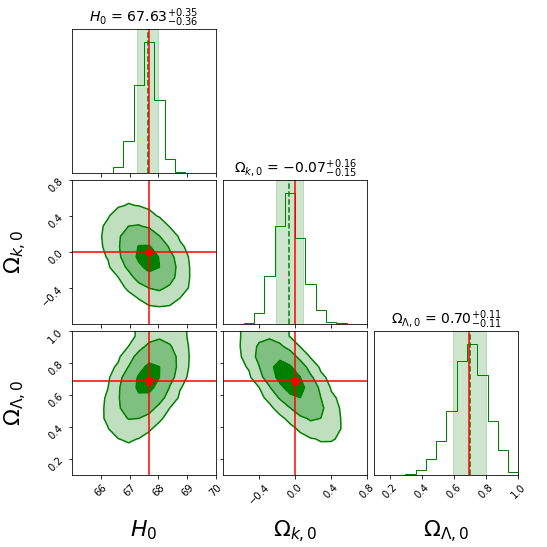}
    \includegraphics[width=0.32\textwidth]{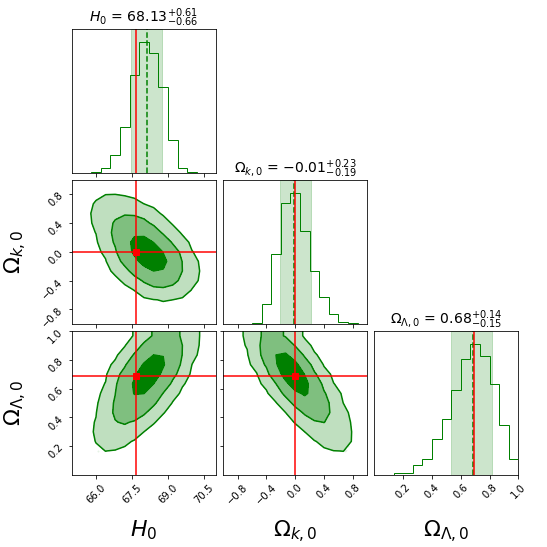}\\
    \includegraphics[width=0.33\textwidth]{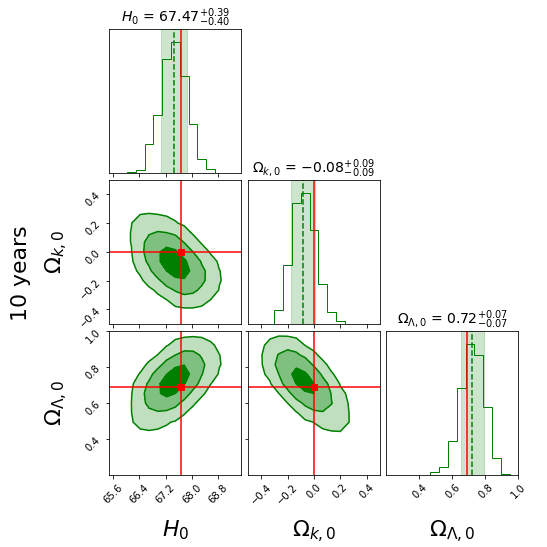}
    \includegraphics[width=0.33\textwidth]{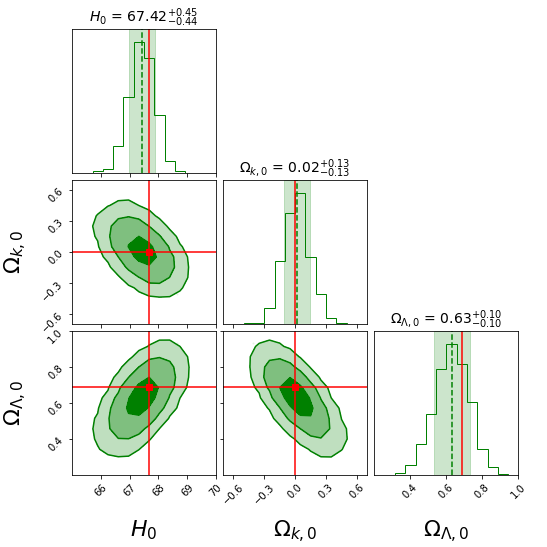}
    \includegraphics[width=0.32\textwidth]{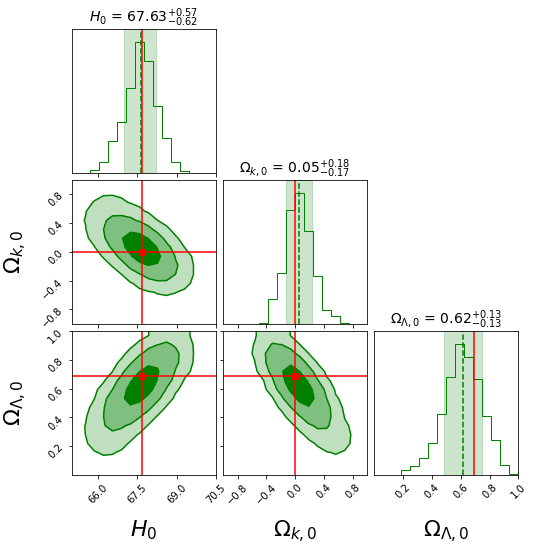}
    \caption{The same of Figure~\ref{fig:contours_realistic_CASE_I} but for the optimistic analysis.}
    \label{fig:contours_optimistic_CASE_I}
\end{figure}

\clearpage
\newpage
\section{Contours plots relative to the analysis including the selection effects and carried out on the GW events with a detected electromagnetic counterpart }\label{App:B}
\begin{figure}[H]
    \centering
    \includegraphics[width=0.33\textwidth]{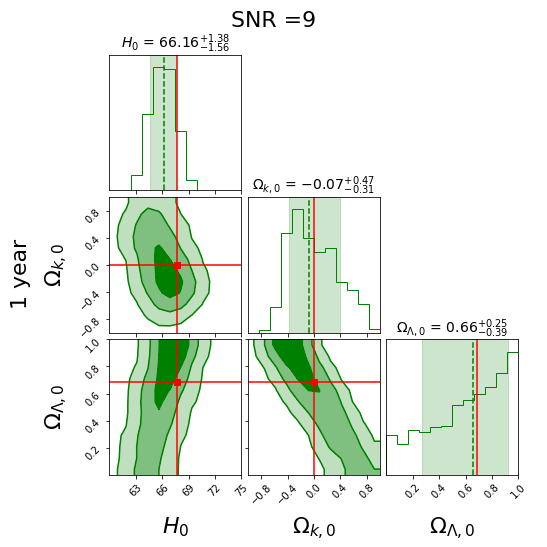}
    \includegraphics[width=0.33\textwidth]{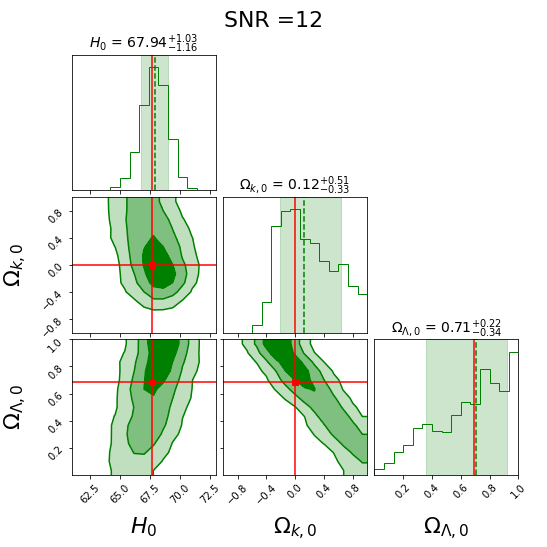}
    \includegraphics[width=0.32\textwidth]{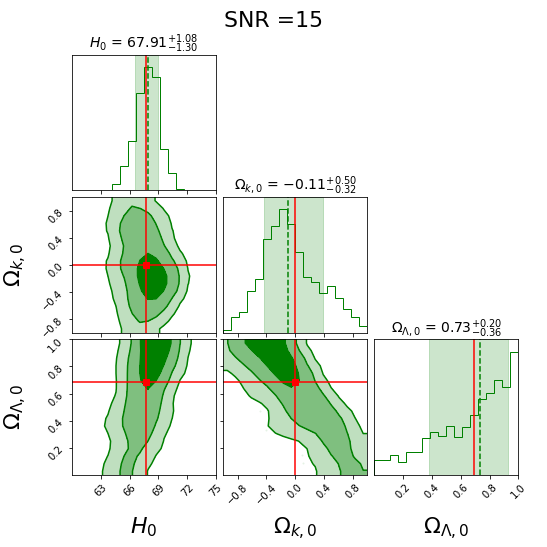}\\
    \includegraphics[width=0.33\textwidth]{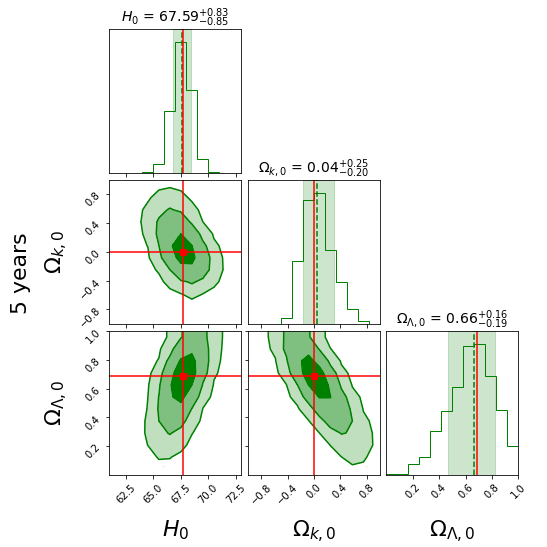}
    \includegraphics[width=0.33\textwidth]{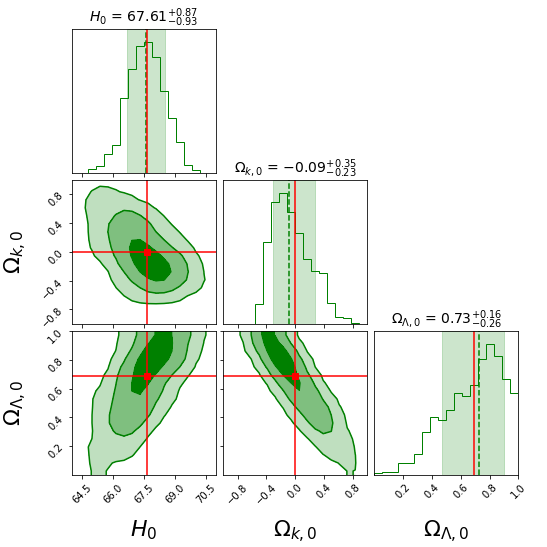}
    \includegraphics[width=0.32\textwidth]{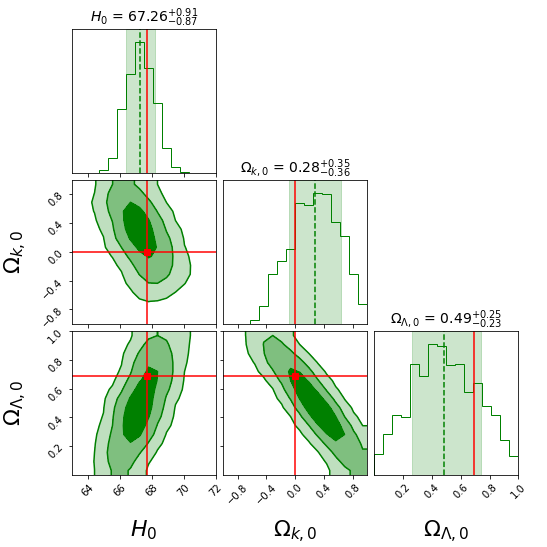}\\
    \includegraphics[width=0.33\textwidth]{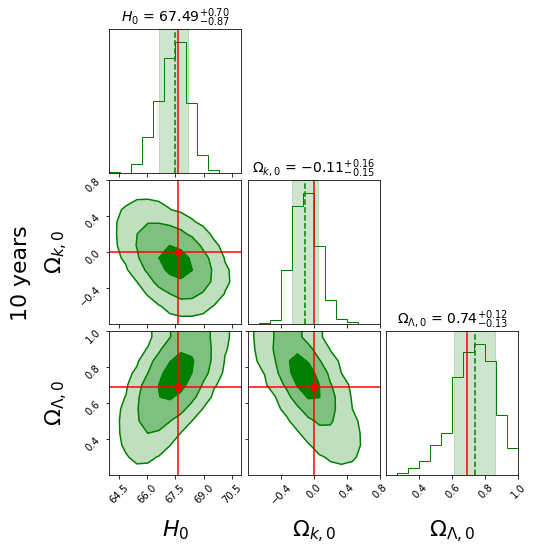}
    \includegraphics[width=0.33\textwidth]{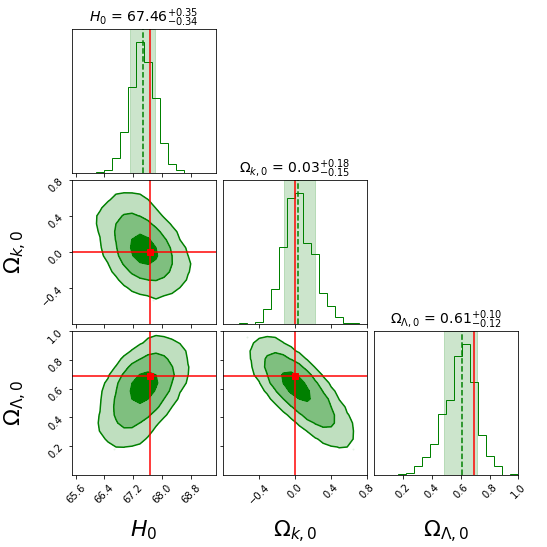}
    \includegraphics[width=0.32\textwidth]{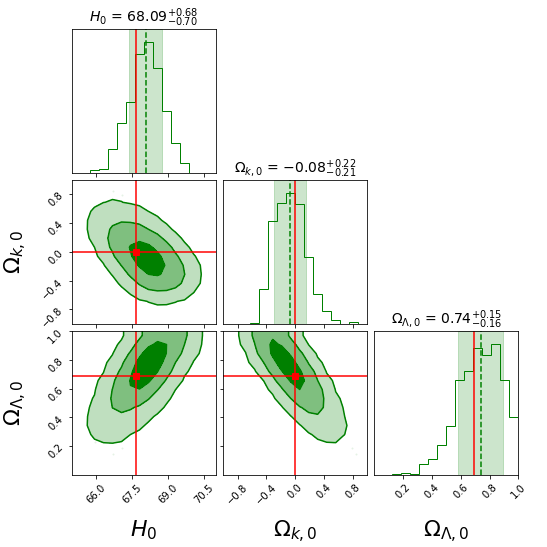}
    \caption{The same of Figure~\ref{fig:contours_realistic_CASE_I} but including the selection effects as discussed in Section~4.}
\end{figure}
\begin{figure}
    \centering
     \includegraphics[width=0.33\textwidth]{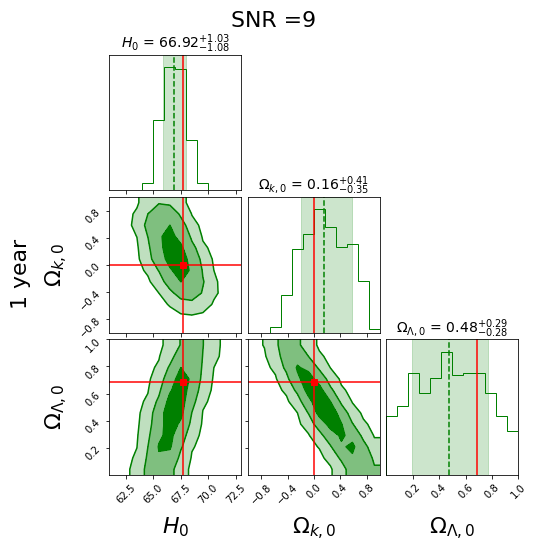}
    \includegraphics[width=0.33\textwidth]{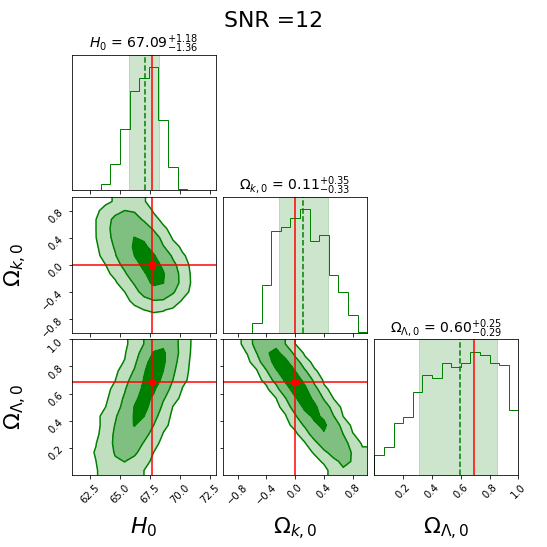}
    \includegraphics[width=0.32\textwidth]{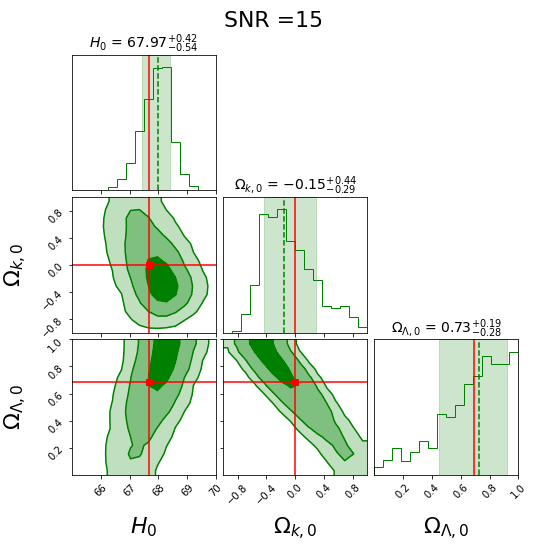}\\
    \includegraphics[width=0.33\textwidth]{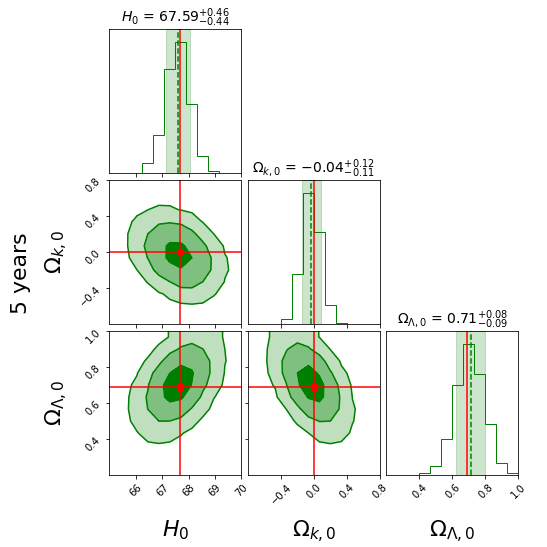}
    \includegraphics[width=0.33\textwidth]{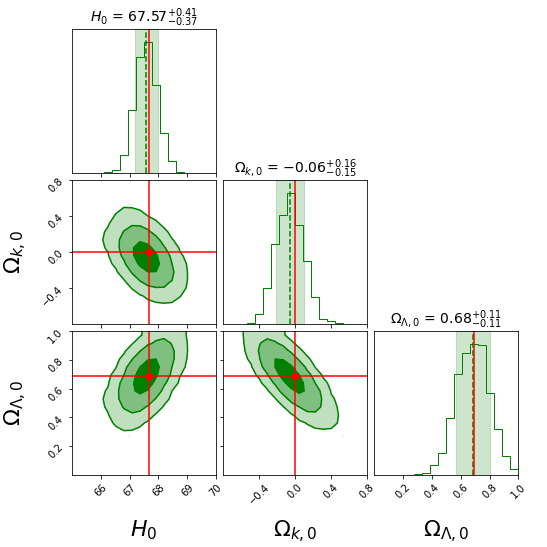}
    \includegraphics[width=0.32\textwidth]{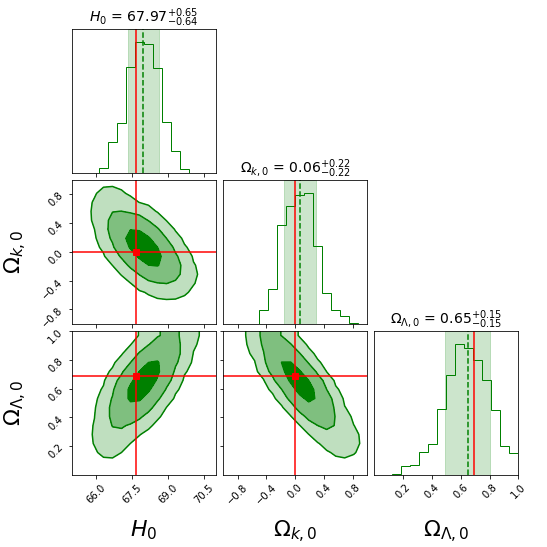}\\
    \includegraphics[width=0.33\textwidth]{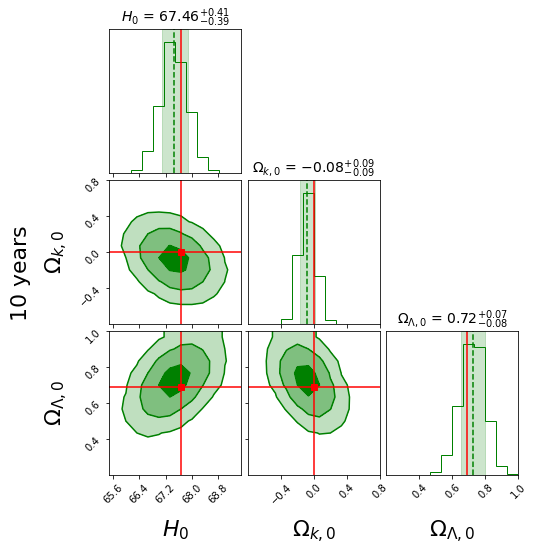}
    \includegraphics[width=0.33\textwidth]{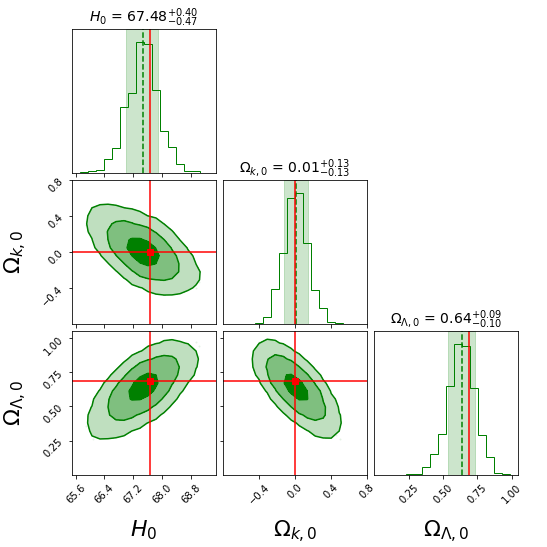}
    \includegraphics[width=0.32\textwidth]{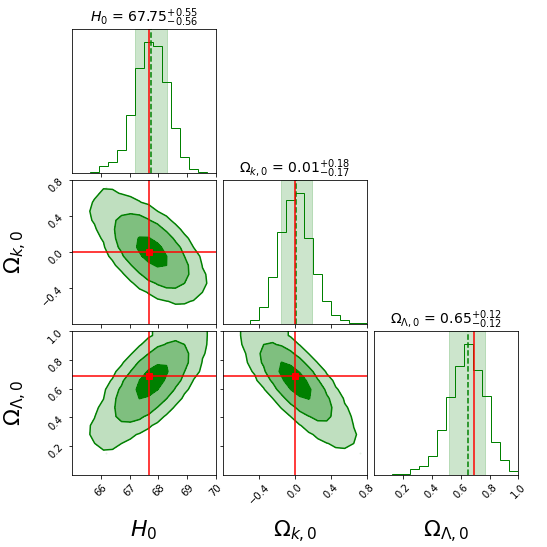}
    \caption{The same of Figure~\ref{fig:contours_realistic_CASE_I} but for the optimistic analysis and including the selection effects as discussed in Section~4.}
\end{figure}

\clearpage
\newpage
\section{Contours plots relative to the analysis  on the dark sirens assuming a prior redshift distribution}\label{App:C}
\begin{figure}[H]
    \centering
    \includegraphics[width=0.33\textwidth]{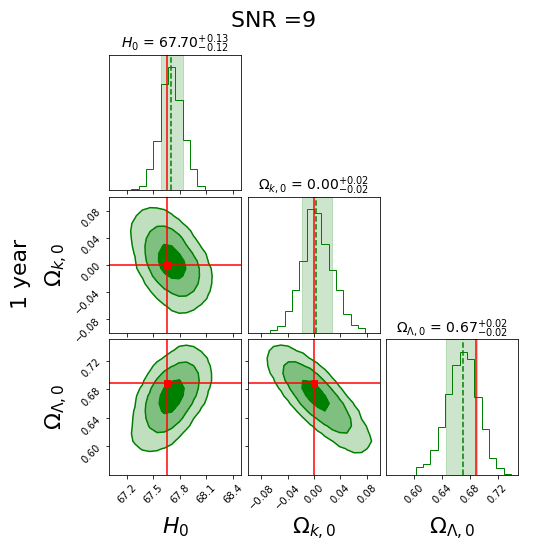}
    \includegraphics[width=0.33\textwidth]{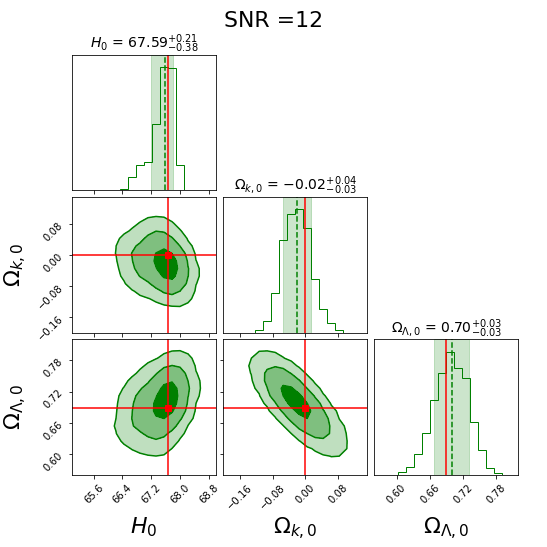}
    \includegraphics[width=0.32\textwidth]{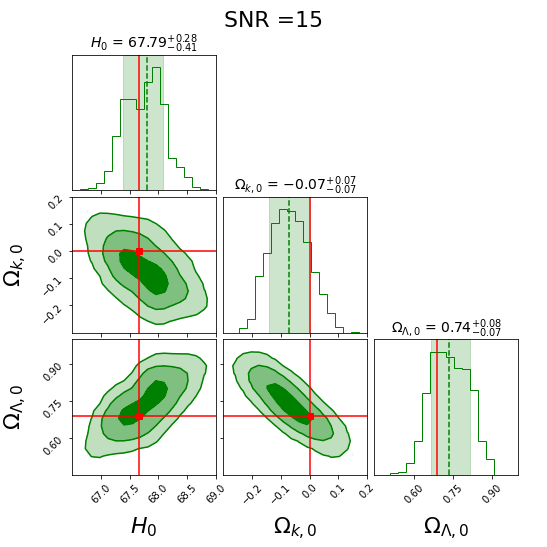}\\
    \includegraphics[width=0.33\textwidth]{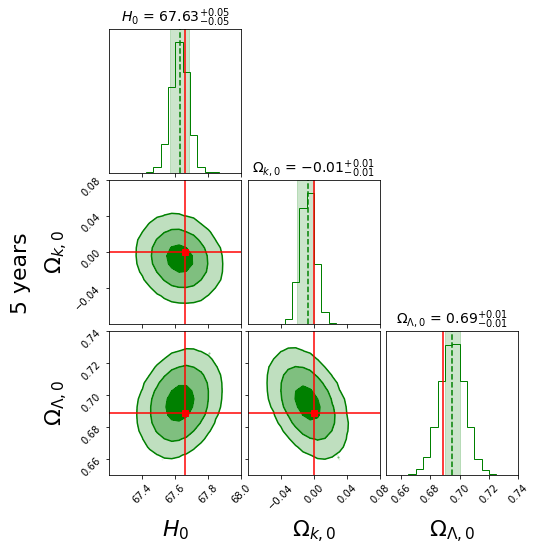}
    \includegraphics[width=0.33\textwidth]{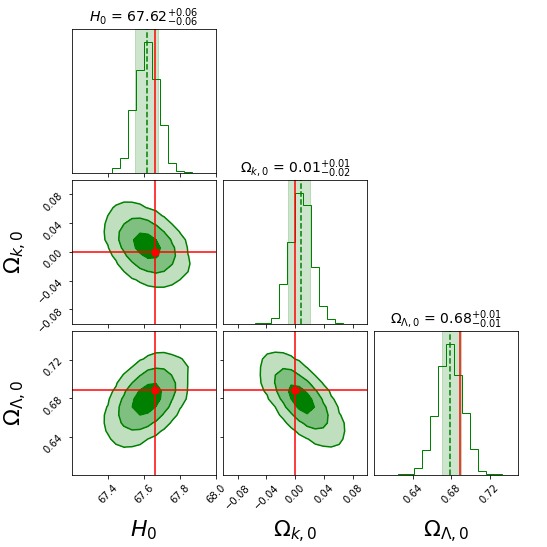}
    \includegraphics[width=0.32\textwidth]{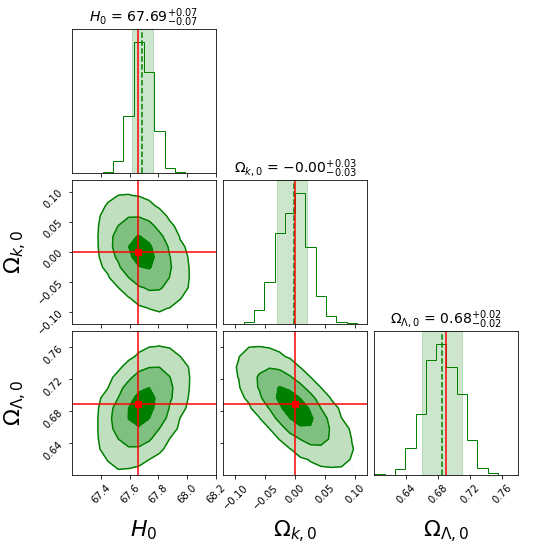}\\
    \includegraphics[width=0.33\textwidth]{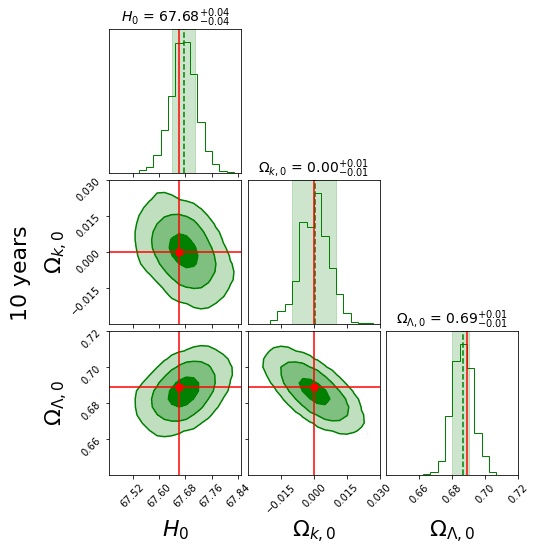}
    \includegraphics[width=0.33\textwidth]{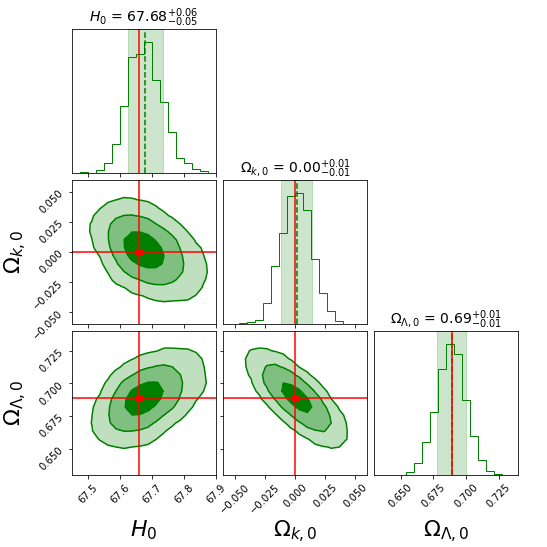}
    \includegraphics[width=0.32\textwidth]{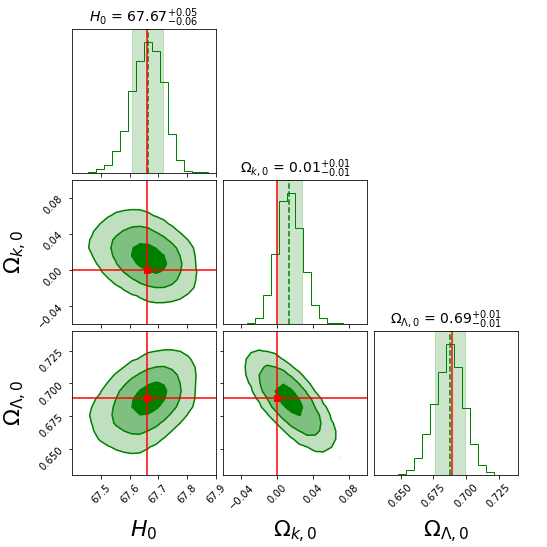}
    \caption{The same of Figure~\ref{fig:contours_realistic_CASE_I} but for the dark sirens assuming known the redshift prior distribution}
    \end{figure}

	\bsp	
	\label{lastpage}
\end{document}